\documentclass[useAMs,a4paper]{mn2e}
\usepackage{savesym}
\usepackage{graphicx}
\expandafter\let\csname equation*\endcsname\relax
  \expandafter\let\csname endequation*\endcsname\relax 
\usepackage{subfig}
\usepackage{amsmath}
\usepackage{amssymb}
\usepackage{verbatim}
\usepackage[yyyymmdd,hhmmss]{datetime}
\usepackage{array}
\usepackage{times}
\usepackage[total={17.8cm,24.0cm},centering]{geometry} 
\usepackage{color}

\newcommand{\be}{\begin{equation}}
\newcommand{\beq}{\begin{equation}}
\newcommand{\ee}{\end{equation}}
\newcommand{\eeq}{\end{equation}}
\newcommand{\eea}{\end{eqnarray}}
\newcommand{\bea}{\begin{eqnarray}}

\newcommand{\dd}{\partial}

\newcommand\W {{W^r_{\ \phi}}}
\newcommand\T {{\widetilde {T}}}

\title[Upper BH mass limit of thermal X-ray TDEs]{An upper observable black hole mass scale for tidal disruption events with thermal X-ray spectra}%{The X-ray properties of disc-dominated tidal disruption events \\
\author [Andrew Mummery, Steven A. Balbus]{Andrew Mummery\thanks{E-mail:
andrew.mummery@physics.ox.ac.uk}, {Steven A. Balbus}
\\
Oxford Astrophysics, Denys Wilkinson Building, Keble Road, Oxford, OX1 3RH, United Kingdom}
\begin{document}

\date{}

\pagerange{\pageref{firstpage}--\pageref{lastpage}} \pubyear{2021}

\maketitle

\label{firstpage}

\begin{abstract} 
We comprehensively model the X-ray luminosity emergent from time dependent relativistic accretion discs, developing analytical models of the X-ray luminosity of thermal disc systems as a function of black hole mass $M$, disc mass $M_d$, and disc $\alpha$-parameter. The X-ray properties of these solutions will be directly relevant for understanding TDE observations. 
We demonstrate an extremely strong suppression of thermal X-ray luminosity from large mass black holes, $L_X \sim \exp(-m^{7/6})$, where $m$ is a dimensionless mass, roughly the the black hole mass in unity of $10^6$M$_\odot$. This strong suppression results in upper-observable black hole mass limits, which we demonstrate to be of order $M_{\rm lim} \simeq 3 \times 10^7 M_\odot$, above which thermal X-ray emission will not be observable. This upper observable black hole mass limit is a function of the remaining disc parameters, and the full dependence can be described analytically (eq. \ref{maxmass_schwarz}). We demonstrate that the current population of observed X-ray TDEs is indeed consistent with an upper black hole mass limit of order $M \sim 10^7M_\odot$, consistent with our analysis.
\end{abstract}


\noindent
%Complied at \today\ \currenttime\ .

\section{Introduction}
There is good observational and theoretical evidence that the dominant emission components of many tidal disruption event (TDE) light curves come from evolving thin discs  (van Velzen {\it et al.} 2019, Mummery \& Balbus 2020a, b).   Late time UV observations of six well-observed TDEs show transitions to disc-dominated states after $\sim 100$ days (van Velzen {\it et al.} 2019), and the spectra of a large sample of X-ray TDEs show properties which are analogous to the high-luminosity spectrally-soft accretion state of X-ray binaries (e.g., Saxton {\it et al}. 2012; Miller {\it et al}. 2015; Lin {\it et al}. 2015; Holoien {\it et al}. 2016a; Gezari {\it et al}. 2017; Wevers {\it et al}. 2019b; Jonker {\it et al}. 2020, Wen {\it et al}. 2020). 

Motivated by these findings, the authors recently developed and applied time-dependent relativistic thin disc theory (Balbus 2017, Balbus \& Mummery 2018) to the light curves of the particularly well-observed TDE ASASSN-14li (Holoein et al 2016a, Mummery \& Balbus 2020a).   In comparison to other approaches, the full $\sim 900$ days of X-ray observations are very well-described by a disc model.    Moreover, the disc model simultaneously, and very naturally, fit the final $\sim 1200$ days of UV observations in three different wavebands. 

The observational evidence suggests that ASASSN-14li is a member of a relatively common class of X-ray TDEs with extremely soft X-ray spectra.   In this and two companion papers, we carry out detailed analytical and numerical studies of the more general X-ray properties of the solutions of the relativistic disc equations, and compare these findings with observed X-ray TDEs. In this paper we shall focus on the properties of the X-ray light curves from discs whose bolometric luminosity is sub-Eddington. In companion papers, we examine disc solutions in the super-Eddington and low-hard spectral state regimes. 

We extend the asymptotic expansion techniques developed in Mummery \& Balbus (2020a), applying them to the high energy disc spectrum.    We find sensitive dependencies of the X-ray luminosity on five key parameters: the black hole mass $M$, the black hole spin $a$, the disc mass $M_d$, the disc-observer inclination angle $\theta$, and the Shakura-Sunyaev $\alpha$ parameter.  These are very general results and should be widely applicable.  The analytic results are confirmed by direct numerical integration of the disc equations.

 Our key result in the sub-Eddington regime is that for a standard $\alpha$-model, the disc X-ray luminosity  is sharply cut-off for large mass black holes.    More precisely,   the X-ray luminosity $L_X$ scales as
\beq
L_X \propto m^{n} \exp\left(-m^{7/6}\right) , 
\eeq
where $n = -1/3$ and $n = 1/4$ for finite and vanishing ISCO stresses respectively, and $m$ is a dimensionless, normalised mass variable, in essence the black hole mass in units of $\sim 10^6$M$_\odot$. 

This paper represents the first part of a four paper TDE unification scheme (Mummery \& Balbus 2021b, Mummery 2021a, b). The layout of the paper is as follows. The technical derivation of the analytical results used throughout the papers is presented in sections \ref{deriv} and \ref{scalings}. Section \ref{num1} then sets up the equivalent numerical problem for a Schwarzschild black hole, and demonstrates that the analytical results of the previous sections reproduce the fully numerical results. Observational implications of the numerical and analytical results are discussed in section \ref{imp}.  The effects of the properties of the ISCO stress are explored in section \ref{vansec}, before the effects of black hole spin and disc-observer inclination angle are explored in section \ref{spinsec}. A comparison to the historic TDE population is performed in section \ref{compsec}. We briefly discuss how the low cadence of wide-field surveys may effect our results in section \ref{survey}, before concluding in section \ref{conc}. 

\section{Analysis}\label{deriv}
In this section, we present a technical derivation of the analytical results used throughout this paper and its companions.  The derivation follows Balbus (2014) in applying Laplace expansion techniques (Bender \& Orszag 1978) to the large photon energy disc spectrum, and is an extension of earlier results presented in Mummery \& Balbus (2020a).  The extension involves computing higher-order correction terms to the earlier leading-order results.  These terms are important for very luminous X-ray sources. 
 
The practical reader mainly concerned with the implications of the analysis for the observed properties of tidal disruption events may wish to skip directly to section \ref{num1}.  In sections \ref{num1}, \ref{vannum}, \ref{spinsec} and \ref{survey} we present a numerical analysis of the disc solutions as observed at X-ray energies. These numerical solutions suffice in themselves to determine all that is needed to understand the implications of our work for observational purposes.   Furthermore, the conclusions of the analysis can be qualitatively understood by rather simple physical arguments, which are presented in section \ref{imp}.  

\subsection{Spectral integral}
The frequency-specific flux density $F_\nu$ of the disc radiation, as observed by a distant observer  (subscript $o$), is given by
\beq
F_{\nu}(\nu_o) = \int I_\nu (\nu_o) \, \text{d}\Theta_{{o}} .
\eeq 
Here, $\nu_o$ is the photon frequency and $I_\nu(\nu_o)$ the specific intensity,  both measured at the location of the distant observer O.   The differential element of solid angle subtended by the disk contribution on the observer's sky is $\text{d}\Theta_{{o}}$.  (We reserve the $\Omega$ notation for the rotating disc's angular velocity variable.)
Since $I_\nu / \nu ^3$ is a relativistic invariant (e.g. Misner, Thorne, \& Wheeler 1973), we may write
\beq
F_{\nu}(\nu_o) = \int f_\gamma^3 I_\nu (\nu_e) \, \text{d}\Theta_{{o}},
\eeq 
where the frequency ratio factor $f_\gamma$ is the ratio of $\nu_o$ to the emitted local rest frame frequency $\nu_e$ of a photon originating from a disc coordinate $(r, \phi)$:
\begin{equation}\label{redshift}
f_\gamma(r,\phi) \equiv \frac{\nu_{o}}{\nu_e} = {p_\mu U^\mu\ ({\rm O})\over p_\lambda U^\lambda\ ({\rm E})}=\frac{1}{U^{0}} \left[ 1+ \frac{p_{\phi}}{p_0} \Omega \right]^{-1} ,
\end{equation}
where O and E refer to observer and emitter, respectively.    In the final equality, we introduce the standard disc 4-velocity components $U^\mu$, and the angular velocity $\Omega$ of the disc fluid, defined by  
\beq
\Omega \equiv {{\rm d}\phi\over {\rm d}t} = {U^\phi\over U^0}.
\eeq
(We distinguish the proper time interval ${\rm d}\tau$ and the distant observer coordinate time ${\rm d}t$, with $U^0={\rm d}t/{\rm d}\tau$.)   The covariant quantities $p_\phi$ and $-p_0$ (on the far right) correspond to the angular momentum and energy of the {\em emitted photon}.   These are of course constants of motion for a photon propagating through the Kerr metric.  Except for special viewing geometries, these components must in general be found by numerical ray tracing calculations.   

The disc is assumed to be a (colour-corrected) multi-temperature black body, each disc annulus having a temperature $T$ which is found by solving the underlying disc equations.  As we shall model disc solutions which are accreting at near Eddington rates, we incorporate radiative transfer effects via a simple spectral hardening factor $f_{\rm col}$ (Shimura \& Takahara 1995).  The specific intensity of the locally emitted radiation is then given by a modified Planck function 
\beq\label{planck}
I_\nu(\nu_e) = f_{\rm col}^{-4} B_\nu(\nu_e, f_{\rm col}T) \equiv \frac{2h\nu_e^3}{f_{\rm col}^4 c^2} \left[ \exp\left( \frac{h\nu_e}{k_B f_{\rm col} T} \right) - 1\right]^{-1} .
\eeq
For an observer at a large distance $D$ from the source, the differential solid angle into which the radiation is emitted is
\beq
 \text{d}\Theta_{{o}} = \frac{\text{d}b_1\text{d}b_2}{D^2} ,
\eeq 
where $b_1$ and $b_2$ are appropriate photon impact parameters at infinity  (Li \textit{et al}. 2005).  (The impact parameters are usually denoted $\alpha$ and $\beta$.    Here, however, we reserve $\alpha$ for the Shakura-Sunyaev viscosity parameter (1973) and $\beta$ for the inverse temperature.)  

The observed flux from the disc surface $\cal S$ is therefore
\beq\label{flux}
F_\nu(\nu_o) =  \frac{1}{D^2}\iint_{\cal S} {f_\gamma^3 f_{\rm col}^{-4} B_\nu (\nu_o/f_\gamma, f_{\rm col} T)} ~\text{d}b_1\text{d} b_2.
\eeq
We are interested here in the high energy limit of this expression ($h\nu \gg k_B T$).  An appropriate observational probe of this spectral region is $F_X$, the total X-ray flux observed across a satellite's passband (for the \textit{Swift} telescope for example, this corresponds to photon energies in the range 0.3 to 10 keV).    This is straightforwardly calculated by integrating (\ref{flux}) over the corresponding (observer) frequency range:
\beq\label{FXdef}
F_X = \frac{1}{D^2} \int_{\nu_l}^{\nu_u}\iint_{\cal S} {f_\gamma^3 f_{\rm col}^{-4} B_\nu (\nu_o/f_\gamma , f_{\rm col} T)} ~ {\text{d}b_1\text{d} b_2} \, \text{d}\nu_o ,
\eeq 
where $\nu_l$ and $\nu_u$ correspond to the formal lower and upper frequencies of the satellite's pass-band respectively.

\subsection{High energy spectrum}
We begin by recasting the equations (\ref{flux}) and  (\ref{FXdef}) into a more familiar form, defining an ``effective temperature'' $\T$ by
\beq\label{Teff}
\T(b_1, b_2, t) = f_\gamma(b_1, b_2) f_{\rm col}(b_1, b_2, t) \, T\big( b_1,b_2,t \big) ,
\eeq
where the disc temperature depends implicitly on $b_1$ and $b_2$ through its radial $r$-dependence, and the colour-correction factor will generally depend on the local disc temperature  $T$.    (Note that for an evolving disc, the effective temperature is a time-dependent quantity.)
We define the dimensionless parameter $\Lambda_o$ by
\beq
\Lambda_o \equiv \frac{h \nu_o}{k_B \T_p} ,
\eeq
where $\T_p$ is the maximum effective disc temperature $\T$.  
For the remainder of this paper we shall be interested in the high energy limit, which corresponds to 
\beq
\Lambda_o \gg 1 .
\eeq
In this regime, the observed flux is well-approximated by a modified Wien-tail form
\beq\label{flux2}
F_\nu(\nu_o) = \frac{2h\nu_o^3}{D^2c^2}  \iint_{{\cal S}} f_{\rm col}^{-4} \exp\left( - \frac{h\nu_o}{k_B \T} \right)  ~ {\text{d}b_1  \text{d} b_2} \, .
\eeq 
To make analytic progress, we shall further assume that the disc is observed nearly face-on, and that the colour-correction factor $f_{\rm col}$ in the innermost disc regions is independent of radius.  { Generally, the colour correction factor  increases with increasing disc temperature, as, for a given disc density, a discs absorption opacity is lower at a higher temperatures (Davies {\it et al}. 2006, Done {\it et al}. 2012). However, the maximum amplitude $f_{\rm col}$ can reach in an accretion disc is capped, with its value saturating at high temperatures. This saturation value is set physically by Compton downscattering in the disc, and approximately limits the colour correction factor to the following value 
\beq
f_{\rm col} \sim \left({72 \, {\rm keV} \over k_B T}\right)^{1/9}, \quad T > 10^5\, {\rm K} .
\eeq
For the purposes of computing the integral in equation \ref{flux2} we are interested only in the colour correction factor of the very hottest disc regions, as these are the only disc regions which contribute to the Wien tail flux. As these disc regions are likely to have saturated the colour correction factor (for typical disc parameters they reach temperatures $T > 10^5 {\rm K}$), $f_{\rm col}$ will only depend very weakly on disc temperature, and even more weakly on disc radius, and so it is a reasonable assumption to treat it as a constant in integral \ref{flux2}. Furthermore, at later points in this paper we shall numerically model the colour correction factor within the disc using the full model of Done {\it et al}. (2012). We will demonstrate that a complex model of $f_{\rm col}$ has only a small effect on the {\it X-ray} flux of our disc models (although this is not true for all observing frequencies).
}

While we compute detailed numerical results for general inclination angles below, the face-on analytical solutions are extremely useful for understanding important gross  properties of the more general solutions.   In this face-on limit, we may ignore the effects of relativistic Doppler-boosting of the observed radiation, meaning that the frequency ratio factor $f_\gamma$ is symmetric in the image plane. The above integrals can therefore be expressed in terms of a radial image coordinate $R$
\beq
R \equiv \sqrt{b_1^2 + b_2^2} \, ,
\eeq 
and therefore the flux integral is given by
\beq\label{flux3}
F_\nu(\nu_o) = \frac{4\pi h\nu_o^3}{D^2c^2 f_{\rm col}^4} \int_{R_p}^{\infty} R \exp\left( - \frac{h\nu_o}{k_B \T} \right) {\text{d}R} \, ,
\eeq 
where $R_p$ is the image plane coordinate of the inner disc edge. Note that, due to gravitational lensing of the emitted photons, this image plane inner disc coordinate is not equal to the ISCO radius $R_p \neq R_I$.  Only the hottest parts of the disc very near the temperature maximum will contribute to the observed flux. This integral may therefore be asymptotically expanded via Laplace's method.    Start by Taylor expanding the inverse of the effective temperature about the effective temperature maximum:
\beq\label{betaexpansion}
\beta \equiv (k_B \T)^{-1} = \beta_p + \frac{\partial \beta}{\partial R}  \, (R - R_p) + \sum\limits_{n=2}^{\infty} \beta^{(n)}  \, \frac{(R - R_p)^n}{n!}, 
\eeq
where each derivative is evaluated at $R = R_p$, $\beta_p \equiv (k_B \T_p)^{-1}$, and 
\beq
\beta^{(n)} \equiv \left(\frac{\partial}{\partial R} \right) ^n  \beta .
\eeq
In this section, we assume that the temperature maximum occurs at the inner disc edge, which is appropriate for a disc with a finite ISCO stress.   (Vanishing stress solutions will be presented {in \S \ref{vansec}}.)   Introducing the dimensionless variables 
\beq\label{kdef}
k_n \equiv \beta^{(n)} R_p^n / \beta_p ,
\eeq
and 
\beq
y \equiv (R-R_p)/R_p ,
\eeq
we have
\beq\label{flux5}
F_\nu  = \frac{4\pi h\nu_o^3}{c^2f_{\rm col}^4}\left( \frac{R_p}{D} \right)^2  I(\Lambda_o) ,
\eeq 
with the function $I(\Lambda_o)$ given by
\beq
I(\Lambda_o) = e^{-\Lambda_o} \int_{0}^{\infty} (1 + y) \, F(y,\Lambda_o) \,   \exp\left( - k_1\Lambda_o y \right) \text{d}y \, ,
\eeq
and
\beq\label{Fdef}
F(y,\Lambda_o) = \exp\left(- \, \Lambda_o \sum\limits_{n=2}^{\infty} k_n  \, \frac{y^n}{n!} \right) \, .
\eeq
The large $\Lambda_o$ solution of $I(\Lambda_o)$ is
%\beq
%I(\Lambda_o) = e^{-\Lambda_o} \sum\limits_{m = 0}^{ \infty} \frac{ F^{(m)}(0,\Lambda_o)}{m! }  \left( \frac{ \Gamma(m + 1) }{ (k_1 \Lambda_o)^{m+1} }  + \frac{  \Gamma(m + 2) }{(k_1 \Lambda_o)^{m+2} } \right) ,
%\eeq
%{\bf ( Do we actually need this formula??   $m$ is always an integer, so just go to the next.)}
 % For integer $m$ this becomes
\beq\label{Isol}
I(\Lambda_o) = e^{-\Lambda_o} \sum\limits_{m = 0}^{ \infty} \frac{  F^{(m)}(0,\Lambda_o)  }{ (k_1 \Lambda_o)^{m+1} } \left( 1  + \frac{ m+1 }{k_1 \Lambda_o } \right) .
\eeq
where $F^{(m)}(0,\Lambda_o)$ is the $m^{\text{th}}$ derivative of $F(y,\Lambda_o)$, evaluated at $y = 0$\footnote{Knowledgeable readers will recognise this as an application of Watson's lemma (Bender \& Orszag 1978).}. 
We have found that over the entire parameter space of interest,
%relevant for modelling the tidal disruption of stars by super massive black holes, 
truncating this sum at order $\Lambda_o^{-3}$ accurately reproduces the properties of the exactly calculated numerical X-ray light curves.    We shall therefore neglect terms of order $\Lambda_o^{-4}$ and higher. To truncate this sum at order $\Lambda_o^{-3}$, we require derivatives $F^{(m)}(y,\Lambda_o)$ up to and including $m = 4$.   These derivatives follow from the definition (\ref{Fdef})
\begin{align}
F^{(0)}(0,\Lambda_o) &= 1, \\
F^{(1)}(0,\Lambda_o) &= 0, \\
F^{(2)}(0,\Lambda_o) &= -k_2 \Lambda_o  , \\
F^{(3)}(0,\Lambda_o) &= - k_3 \Lambda_o  , \\
F^{(4)}(0,\Lambda_o) &= - k_4\Lambda_o + 3 k_3^2 \Lambda_o^2 .
\end{align}
The flux integral is therefore given by
\beq
I(\Lambda_o) = e^{-\Lambda_o} \left[ \frac{c_1}{ \Lambda_o} + \frac{c_2}{\Lambda_o^2} + \frac{c_3}{\Lambda_o^3}+ O\left(\Lambda_o^{-4} \right) + ...\, \right] ,
\eeq
where 
\begin{align}
c_1 &= 1/k_1, \\
c_2 &= (k_1 - k_2)/k_1^3 , \\
c_3 &=  (3k_3^2 -3k_1k_2 - k_1k_3)/k_1^5 .
\end{align}

To gain a sense of the scale of the numerical values of the $c$-coefficients, consider a simple disc temperature profile 
\beq\label{simpT}
T \propto R^{-q} ,\quad q > 0,
\eeq
and neglect the effects of gravitational red-shift.   In this limit 
\beq
k_n = {\prod\limits_{j = 1}^n}  \, (q - j + 1) ,
\eeq 
and
\begin{align}
c_1 &= 1/q  , \\
c_2 &= (2-q)/q^2 ,  \\
c_3 &= (3q^4 - 18q^3 + 38q^2 - 36q + 13)/q^3 .
\end{align}
These coefficients are all positive for $0<q < 1$.  For reference, classical disc models are well approximated by equation (\ref{simpT}) in their inner regions, with $q = 3/4$ (vanishing ISCO stress) or $q = 7/8$ (finite ISCO stress{; Mummery \& Balbus 2020a}). For the case of finite ISCO stress, we find
\begin{align}
c_1 &= 8/7 \approx 1.14 , \\
c_2 &= 72/49 \approx  1.47 , \\
c_3 &= 1203/2744 \approx 0.44 .
\end{align}

\subsection{X-ray flux}\label{analyticalflux}
With a functional form for the spectrum in place, we may now calculate the observed X-ray flux:
\beq
F_X = \int_{\nu_l}^{\infty} F_\nu(\nu_o) \, \text{d} \nu_o = (h \beta_p)^{-1} \int_{\Lambda}^{\infty} F_\nu(\Lambda_o) \, \text{d} \Lambda_o,
\eeq
where have defined 
\beq
\Lambda \equiv \frac{h \nu_l}{k_B \T_p} \gg 1 ,
\eeq
and have extended the upper integration limit to infinity, which introduces only exponentially small corrections.  
The X-ray luminosity is then of the form 
\beq
F_X = \frac{4\pi h  \nu_l^4}{c^2 f_{\rm col}^4} \left(\frac{R_p}{D}\right)^{2} \,  K(\Lambda) ,
\eeq
where $K(\Lambda)$ is the expansion:
\beq
K(\Lambda) = \frac{1}{\Lambda^{4}}  \int_{\Lambda}^{\infty}  e^{-\Lambda_o} \left[ c_1 \Lambda_o^2 + c_2 \Lambda_o + c_3  \right] \text{d} \Lambda_o.
\eeq
This becomes
\beq
K(\Lambda) = \frac{ e^{-\Lambda}}{\Lambda^{4}}  \Bigg[ (2 + 2\Lambda + \Lambda^2) c_1 + (1 + \Lambda) c_2 + c_3 \Bigg] .
\eeq
The X-ray luminosity is then given by
\beq\label{FX}
F_X = F_0 \left(\frac{R_p}{D}\right)^{2}  \left[ \frac{1}{\Lambda^2} + \frac{\psi_1}{\Lambda^3} + \frac{ \psi_2}{\Lambda^4} \right] e^{-\Lambda} ,
\eeq
with amplitude 
\beq\label{amp}
F_0 = \frac{4\pi }{c^2 f_{\rm col}^4} h\nu_l^{4}   c_1 .
\eeq
{The factor $(R_ p/D)^2$, is related to the angular size of the disc on the sky,  and the coefficients are:}
\begin{align}
\psi_1 &=2 +  c_2/c_1, \\
\psi_2 &= 2 +  (c_2 + c_3)/c_1 .
\end{align}
For a disc with a temperature  $T \propto R^{-7/8}$, these have numerical values  
\begin{align}
\psi_1 &\simeq 3.29 , \\
\psi_2 &\simeq 3.67  .
\end{align}
Equation (\ref{FX}) is very general, and holds for thermal emission from a finite ISCO stress disc whenever $h \nu_l > k_B T_p$.   Note that {\it only} the temperature of the hottest part of the disc enters this equation.  In this regard, the X-ray properties of these disc solutions are extremely simple to describe:  one only needs to understand the properties of the hottest regions, not the global disc properties.  The dependence of the disc temperature upon system parameters may be determined analytically, once a turbulent stress parameterisation and initial condition are specified. This is done in the following section.

%{\bf Reading through 20 August 15:10}

\section{Temperature, X-ray flux, and bolometric luminosity scalings}\label{scalings}

\subsection{Disc temperature}
The dominant $r$-$\phi$ component of the turbulent stress tensor $\W$ serves to transport angular momentum throughout the disc as well as to extract the free energy of the disc shear.  This free energy is then thermalised and radiated from the disc surface.   Both the extraction and the dissipation are assumed to be local processes.    These standard assumptions lead to a disc temperature profile given, in relativistic theory, by (e.g., Balbus 2017)
\beq
\sigma T^4 =  - \frac{U^0U^\phi}{2r}  ~ (\ln\Omega) '~ \zeta(r,t), 
\label{temperature}
\eeq
where
\beq
 \zeta \equiv  r \Sigma \W/U^0,
\eeq
and $\sigma$ is the standard Stefan-Boltzmann constant.  In terms of more practical physical variables, equation (\ref{temperature}) reads 
\beq\label{temp}
\sigma T^4 = \frac{3\sqrt{GM}}{4\,r^{7/2}} \frac{\zeta(r,t)}{1 - {3r_g}/{r} + 2a\sqrt{{r_g}/{r^3}}} .
\eeq
{ where the gravitation radius $r_g$ is defined by $r_g = GM/c^2$.}

Here, we are interested in the general properties of the solutions of the thin disc equations as observed at X-ray energies.  An interesting and useful observational probe of these solutions is the evolution of the peak flux of the disc's X-ray light-curve.   As argued in the previous section, in the spectral range of interest, the peak flux will be a function of the highest temperatures %{\bf (note the subtle change here, we are looking for a series of high temps)}  
reached in the disc during its overall evolution. 

There are five physical variables which completely describe the characteristic peak temperature scale of the disc:  the initial disc mass $M_d$; the initial disc radius $r_0$; the black hole mass $M$; the black hole spin $a$; and the magnitude of the turbulent stress, which is expressed in terms of the Shakura-Sunyaev parameter $\alpha$.  The peak disc temperature is particularly sensitive to the three parameters $M_d$, $M$ and $\alpha$.   When expressed in terms of these three variables,  radial scales (such as the ISCO) vary linearly with the black hole mass 
\beq\label{rm}
r \propto M , 
\eeq
and by modelling the disc as an initial ring of material laid down at some radius $r_0$, the surface density scales as  
\beq
\Sigma \propto {M_d}/{r_0^2} \propto{M_d}{M^{-2}} .\label{sigma}
\eeq

The form of the turbulent stress cannot be found from first principles of course; some prescription is needed.    As in our previous work (Mummery \& Balbus 2019a) we compute the turbulent stress using a modified version of the standard $\alpha$-disc model of Shakura \& Sunyaev (1973).    Whereas $\alpha$-disc models generally set the dynamical stress proportional to the total pressure of the disc (including, most importantly, that of radiation), here we scale the stress proportionately to just the {\it gas} pressure within the disc.   While {\it ad hoc,} the principal advantage of this approach is that it produces dynamically stable disc models, which otherwise are prone to the Lightman--Eardley (1974) instability in the inner radiation-dominated disc regions.   Alternatively, one could model the stress as an arbitrary but stable bi-power law in $\Sigma$ and $r$.   The precise details of modelling are not critical, as long as evolution is stable.   Our $\alpha$ prescription is just a special case with rather simple physics.   

Mathematical convenience is one thing, actual behaviour is another.    At present, it is not clear to what extent the Lightman--Eardley instability represents a true physical instability.   Local 3D disc simulations with radiative hydro do exhibit unstable behaviour (Jiang, Stone \& Davis 2013), as do global 2D hydro simulations (Fragile et al. 2018); nevertheless, observations of X-ray binaries are completely compatible with stable thermal discs (e.g. Done, Gierlinski \& Kubota 2007).   Perhaps most strikingly, the TDE ASASSN-14li was detected at X-ray energies for almost 1000 days (Bright {\it et al}. 2018).  This light curve was  well-described by thermal emission from a stable accretion disc (Mummery \& Balbus 2020a), despite having an Eddington ratio (at peak brightness) of $\sim 0.85$, which would be formally unstable by way of the usual $\alpha$-disc modelling. 

Finally, the degree to which a particular stress parameterisation accurately models the properties of a true turbulent stress is any case very difficult to quantify.    As noted above, our principal conclusions are not sensitive to the precise functional form of the turbulent stress.   Our approach should capture the essential disc dynamics, provided only that {\it some} well-defined, local enhanced stress tensor exists, even if it is not of the precise mathematical form used here. 

The turbulent stress $\W$ in our simplified $\alpha$-disc model is defined by:
\beq\label{stressdef}
\W \equiv \alpha r c_s^2,
\eeq
where $c_s$ is the isothermal gas sound speed, which is related to the gas pressure $P_g$ and density $\rho$ through the standard ideal gas equation of state
\beq
c_s^2 = {P_g \over \rho} =  \frac{k_B T_c}{\mu m_p} .
\eeq
In this expression, $T_c$ is the disc mid-plane temperature, and $\mu$ is the mean molecular mass of a fluid element in units of the proton mass $m_p$.   The extra factor of $r$ in equation (\ref{stressdef}) is a consequence of its mixed tensorial form:  with a covariant $\phi$ index, $\W$ measures correlated fluctuations between the angular {\it momentum} and radial velocity of the fluid, rather than fluctuations in the ordinary circular and radial velocities of the fluid. 

In the simplest radiative $\alpha$ models, the central mid-plane temperature of the disc $T_c$ is related to the surface temperature $T$ (eq. \ref{temp}) by
\beq
T_c^4 = \frac{3}{8} \kappa \Sigma T^4 ,
\eeq
where $\kappa$ is the disc opacity.   In the Newtonian limit, the surface temperature is given by
\beq\label{newtontemp}
\sigma T^4 = -\frac12 \W \Sigma \Omega' . 
\eeq
The opacity within the disc is expected to be dominated by electron scattering, and  we therefore assume that the total opacity is constant and equal to
\beq\label{opacity}
\kappa = \kappa_{\rm es} \simeq 0.034\, {\rm m}^2 {\rm kg}^{-1} .
\eeq
Results for different opacity laws are presented in Appendix \ref{opacity_params}.  The system of equations (\ref{stressdef}--\ref{opacity}) is closed, with a resulting turbulent stress given by 
\beq\label{alphastress}
\W = C\, {\Sigma}^{2/3} {r}^{1/2}, 
\eeq
where $C$ is dimensional constant 
\beq\label{stressamp}
C \equiv \alpha^{4/3} \left(\frac{k_B}{\mu m_p}\right)^{4/3}  \left[ \frac{9}{32\sigma}\kappa_{\rm es} \sqrt{GM} \right]^{1/3} .
\eeq
Equation (\ref{newtontemp}) then simplifies to
\beq
\sigma T^4 = \frac{3 C}{4r^2}\Sigma^{5/3} \sqrt{GM}  ,
\eeq
and the disc surface temperature $T$ then satisfies the following scaling law:
\beq\label{Tscale}
T \propto \frac{\alpha^{1/3} M_d^{5/12} }{M^{7/6}} .
\eeq
Note that these scalings are thus far not particular to our own modelling; rather, they follow from standard $\alpha$ theory.  Nevertheless,  we shall see that these dependencies have considerable observational significance. 

\subsection{Bolometric luminosity}
Both the peak temperature and the bolometric luminosity of the thin disc equation solutions are  strongly dependent on the model parameters, particularly the black hole mass (eq. \ref{Tscale}).   The luminosity is found by integrating the locally radiated flux over the entirety of the disc.  More explicitly,
\beq\label{bol}
L_{\rm bol}(t) = 2\pi \int_{r_I}^{\infty} ({g_{rr} g_{\phi\phi}})^{1/2} \gamma^\phi(r,a) \, 2 \sigma  T^4(r,t) \, {\rm d}r .
\eeq
Here $\gamma^\phi$ is a relativistic factor relating the disc area element in the rotating disc frame to that of the Boyer-Lindquist coordinate system.    (See Bardeen {\it et al}. 1972 for a detailed discussion.)

The peak bolometric luminosity scaling follows from (\ref{Tscale}):
\beq\label{bolop}
L_{\rm bol, peak} \propto R^2 T_p^4 \propto  \alpha^{4/3} M_d^{5/3} M^{-8/3} .
\eeq
The Eddington luminosity $L_{\rm Edd} \propto M$, hence the luminosity ratio scales as:
\beq\label{edratio}
l \equiv \frac{L_{\rm bol,peak}}{L_{\rm Edd}} \propto \frac{\alpha^{4/3} M_d^{5/3}}{M^{11/3}} \leq 1 .
\eeq
The final inequality is a self-consistency constraint, as $l$ values in excess of unity are unlikely to be compatible with the assumptions of the thin disc model.     The constraints imposed by the Eddington luminosity are important for understanding the X-ray light curves of TDEs. 

\subsection{X-ray flux }

The X-ray flux is very sensitive to the parameter $\Lambda \equiv h\nu_l/k_B\T_p$.   Explicitly displaying the key parameter scalings, whilst absorbing other collatoral factors into an amplitude $A_1$, we have
\beq\label{lamb}
\Lambda = A_1 \frac{M^{7/6}}{ \alpha^{1/3} M_d^{5/12}} .
\eeq
By construction, $A_1$ then depends upon the non-displayed system parameters (notaby the black hole spin $a$), and will be determined numerically. 
Finally, the angular size of the disc on the sky is proportional to $(M/D)^{2}$, and thus the leading order X-ray flux scales as 
\begin{equation}\label{full}
F_X \propto  \frac{1}{D^{2}}  \frac{ \alpha^{2/3} M_d^{5/6}}{M^{1/3}}  \exp\left(-%\Lambda 
A_1\frac{M^{7/6}}{ \alpha^{1/3} M_d^{5/12}}
\right) 
\end{equation}
 This is a key result of this paper.   

It is clear from this analysis that the X-ray flux of evolving relativistic disc solutions will be an extremely sensitive function of the system parameters $M_d$, $\alpha$ and $M$.  The flux depends upon the disc parameters in a more-or-less intuitive manner:  larger mass discs with a greater turbulent stress will be brighter in X-rays than lower mass discs with less turbulence.    Similarly, since discs around more massive black holes penetrate less deeply into the central gravitational potential, they attain lower peak disc temperatures.   For a given disc mass and $\alpha$ parameter,  such cooler discs produce substantially dimmer X-ray light curves around more massive black holes: this is a more important effect than the $M^2$ scaling of the emitting disc area.

\section{numerical evaluation -- fiducial case}\label{num1}
We next { numerically} compute the peak observed flux of the evolving X-ray light curves for a fiducial set of disc parameters.    We consider only those X-ray light curves produced by sub-Eddington discs, computed with the aid of the relativistic disc evolution equation (Balbus 2017), reviewed below. 

\subsection{The relativistic disc equation}
The relativistic disc equation describes the evolution of the azimuthally-averaged, height-integrated disc surface density $\Sigma (r, t)$, using standard cylindrical Boyer-Lindquist coordinates for the Kerr geometry: $r$ (radial), $\phi$ (azimuthal), $z$ (vertical), and $t$ (time).  The contravariant four velocity of the disc fluid is denoted $U^\mu$; its covariant counterpart is $U_\mu$.  The specific angular momentum corresponds to $U_\phi$, a covariant quantity.     We assume that there is an anomalous stress tensor present, $\W$, due to low level disk turbulence.   The stress is a measure of the correlation between the fluctuations in $U^r$ and $U_\phi$ (Balbus 2017), and could also include correlated magnetic fields.  As its notation suggests, $\W$ is a mixed tensor of rank two.    It is convenient to introduce the quantity $\zeta$,
\beq\label{z}
\zeta \equiv \sqrt{g}\Sigma \W / U^0=  r \Sigma \W/U^0,
\eeq
where $g>0$ is the absolute value of the determinant of the (mid-plane) Kerr metric tensor $g_{\mu\nu}$.  The Kerr metric  describes the spacetime external to a black hole of mass $M$ and angular momentum $J$.  For our choice of coordinates, $\sqrt{g}=r$.   The ISCO radius, inside of which the disc is rotationally unstable, is denoted as $r_I$. Other notation is standard: the gravitation radius is  $r_g = GM/c^2$, and the black hole spin parameter $a = J/Mc$. 

Under these assumptions, the governing equation for the evolution of the disc is quite generally given by (Eardley \& Lightman 1975; Balbus 2017):
\beq\label{fund}
{\dd \zeta\over \dd t} =  \mathcal{W} {\dd\ \over \dd r}\left({U^0\over U'_\phi}    {\dd \zeta \over \dd r} \right).
\eeq
where the primed notation denotes a radial gradient, and we have defined the stress-like quantity
\beq
\mathcal{W} \equiv  {1 \over (U^0)^2} \left(\W + \Sigma {\dd\W\over \dd \Sigma} \right)  .
\eeq 
This equation (\ref{fund}) is very similar in overall form to the classic Newtonian disc evolution equation (Lynden-Bell \& Pringle 1974), the primary differences being the time dilation effects embodied in $U^0$, and the differing functional form for the angular momentum gradient $U'_\phi$.  The latter leads to the existence of an ISCO in the relativistic case.   

\subsection{Fiducial model light curve}

\begin{table}
\renewcommand{\arraystretch}{2}
\centering
\begin{tabular}{|p{2.2cm}|p{2.cm}|p{2cm} |}\hline
Parameter  & Value \\ \hline\hline
$a/r_g$ & $0$  \\ \hline
$R_0$ & $30 r_g$ \\ \hline
$\alpha$ & $0.1$ \\ \hline
$M_d$ & $0.5 M_\odot$ \\ \hline
$\theta_{\rm obs}$ & $0^\circ$ \\ \hline
$f_{\rm col}$ &$2$ \\ \hline
$D$ & $100$ Mpc \\ \hline
\end{tabular}
\caption{The parameters used to calculate the fiducial X-ray light curves of Figure \ref{naivemass}.   }
\label{paramtable}
\end{table}

We next compute the $0.3$--$10$ keV X-ray light curves of a simple disc model, with parameters summarised in Table \ref{paramtable}.  The initial disc is described by a numerical delta function ring located at a radius $R_0 = 30r_g$, with initial mass $M_d = 0.5 M_{\odot}$.  The disc is then evolved forward in time using the evolution equation (\ref{fund}), with turbulent stress given by equations (\ref{alphastress} \& \ref{stressamp}) with $\alpha = 0.1$. The numerical simulations in this section all assume a finite ISCO stress. Finite ISCO stress models provide a better fit to the observed light curves of the TDE ASASSN-14li than vanishing-stress models.  To avoid unsustainable behaviour at large times in the form of a divergent luminosity, we follow the `quasi-circular orbit' approach developed in Mummery \& Balbus (2019b), with a $\gamma$ parameter of $\gamma = 0.1$.   This allows a small but finite departure from strictly circular orbits.   (See Mummery \& Balbus (2019b) for further details.)   By way of comparison, results for a vanishing ISCO stress are presented in \S \ref{vansec}. 

The time-dependent disc temperature profile is calculated with equation (\ref{temp}), and photon ray tracing calculations, assuming a face-on disc orientation, are used to compute the discs evolving X-ray light curve (see appendix A of Mummery \& Balbus 2020a for a description of the ray tracing algorithm). { In this section w}e assume that the spectral hardening factor { has saturated in the innermost disc regions, to a value of} $f_{\rm col} = 2${. This value is} in keeping with the analytic estimate of 
\beq
f_{\rm col} \simeq \left(\frac{72\, {\rm keV}}{k_B T}\right)^{1/9}
\eeq
for the typical peak disc energies $k_B T_{\rm max} \sim 100 {\textendash} 150 \, {\rm eV}$ { considered in this work (Davis {\it et al.} 2006, Done {\it et al}. 2012). In the much cooler disc regions at larger disc radii a constant $f_{\rm col}$  will be a poor assumption, however these regions will not contribute to the X-ray flux.  In later sections we will relax this assumption and use the full colour correction model of Done {\it et al}. (2012).}

\begin{figure}
  \includegraphics[width=.5\textwidth]{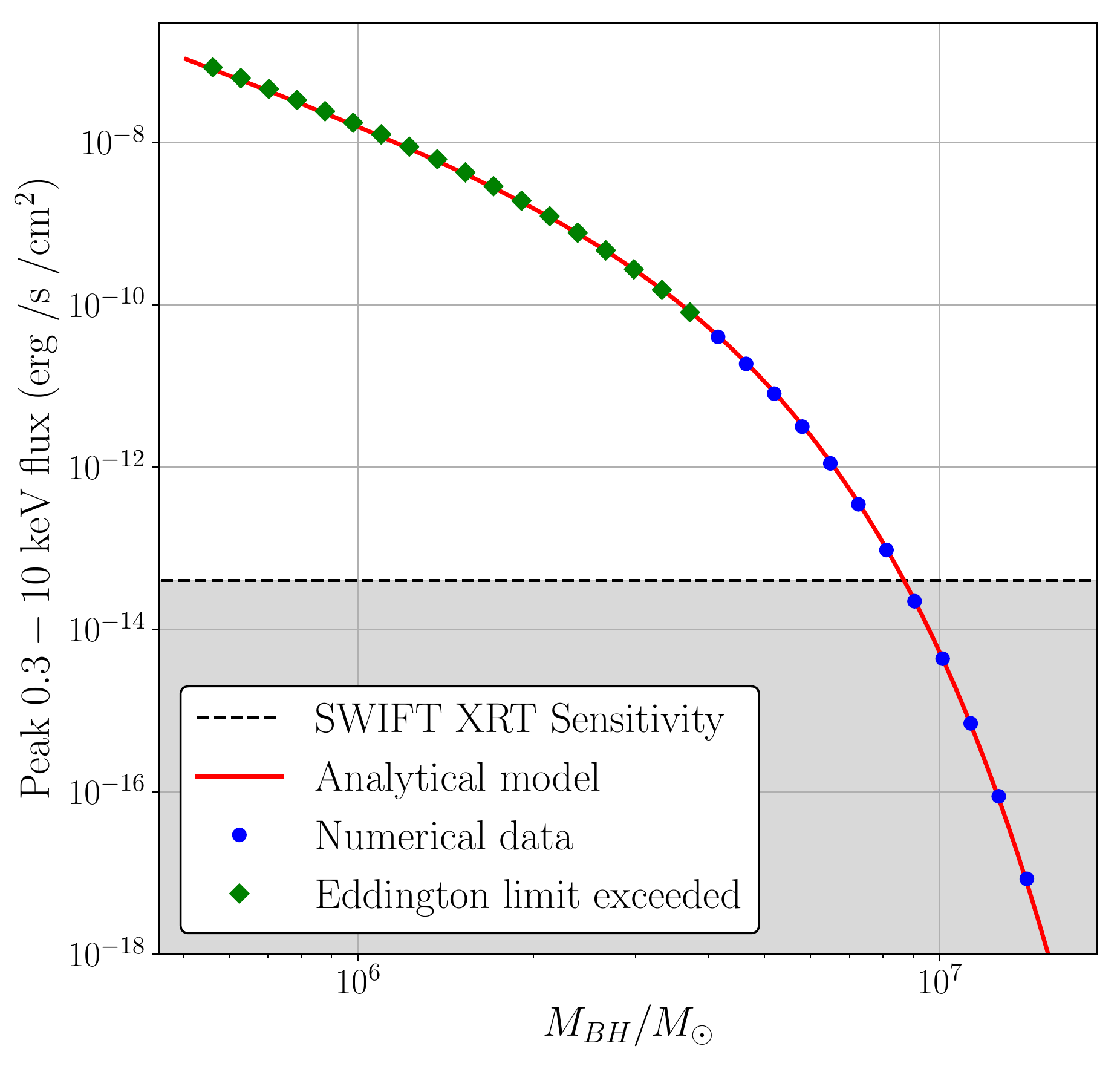} 
 \caption{The peak $0.3$--$10$ keV X-ray flux, assuming a distance of $100$Mpc, for models with an initial disc mass $M_d = 0.5 M_{\odot}$ and $\alpha=0.1$, as a funtion of black hole mass.  Blue dotted points correspond to discs with sub-Eddington peak bolometric luminosities, whereas green diamonds correspond to discs with super-Eddington bolometric luminosities.  The red solid curve is a comparison fit to the theoretical curve eq. [\ref{full}], indistinguishable from the numerical result.  The grey zone lies below {\em Swift} detectability. } 
 \label{naivemass}
\end{figure}

The peak observed { X-ray} flux, assuming a { source-observer} distance { of} $D = 100\,{\rm Mpc}$, is { recorded for each disc solution}.  We repeat this procedure for black hole masses varying between $ M = 5\times10^5 M_{\odot}$ and $M = 1.5 \times 10^7 M_{\odot}$.  The peak observed flux versus black hole mass is plotted as blue points in figure \ref{naivemass}.  The black dashed line corresponds to the {\it Swift} XRTs minimum flux sensitivity\footnote{http://swift.gsfc.nasa.gov/about\_swift/xrt\_desc.html}, $F_{X,{\rm lim}} = 4\times10^{-14} \,{\rm erg}{\rm\, s^{-1}} {\rm cm}^{-2}$. %   {\bf (Avoid long subscripts if at all possible!}) 

We also compute the evolving discs bolometric luminosity using eq. (\ref{bol}).  For each black hole mass, the peak value of the disc bolometric light curve is computed and compared with the Eddington luminosity.  Discs for which this peak value exceeds the Eddington luminosity are plotted with green diamonds.  The governing assumptions of the relativistic disc model will not be valid in this regime, and these values should thus be viewed as purely formal.

We have fit the parameters $F_0, \psi_1, \psi_2$ and the amplitude of the $\Lambda$ parameter $A_1$ to these numerically calculated values. The best fit analytic curve is plotted as the red solid curve.  The analytic expression (eq. \ref{full}) provides an excellent fit to the numerically calculated values.  In Appendix \ref{param_vals} we present the fitting equations in full, in order to facilitate the reproduction of our results.

%{\bf Reading 21 August.}

\subsection{Varying $\alpha$ and $M_d$}
\begin{figure}
  \includegraphics[width=.5\textwidth]{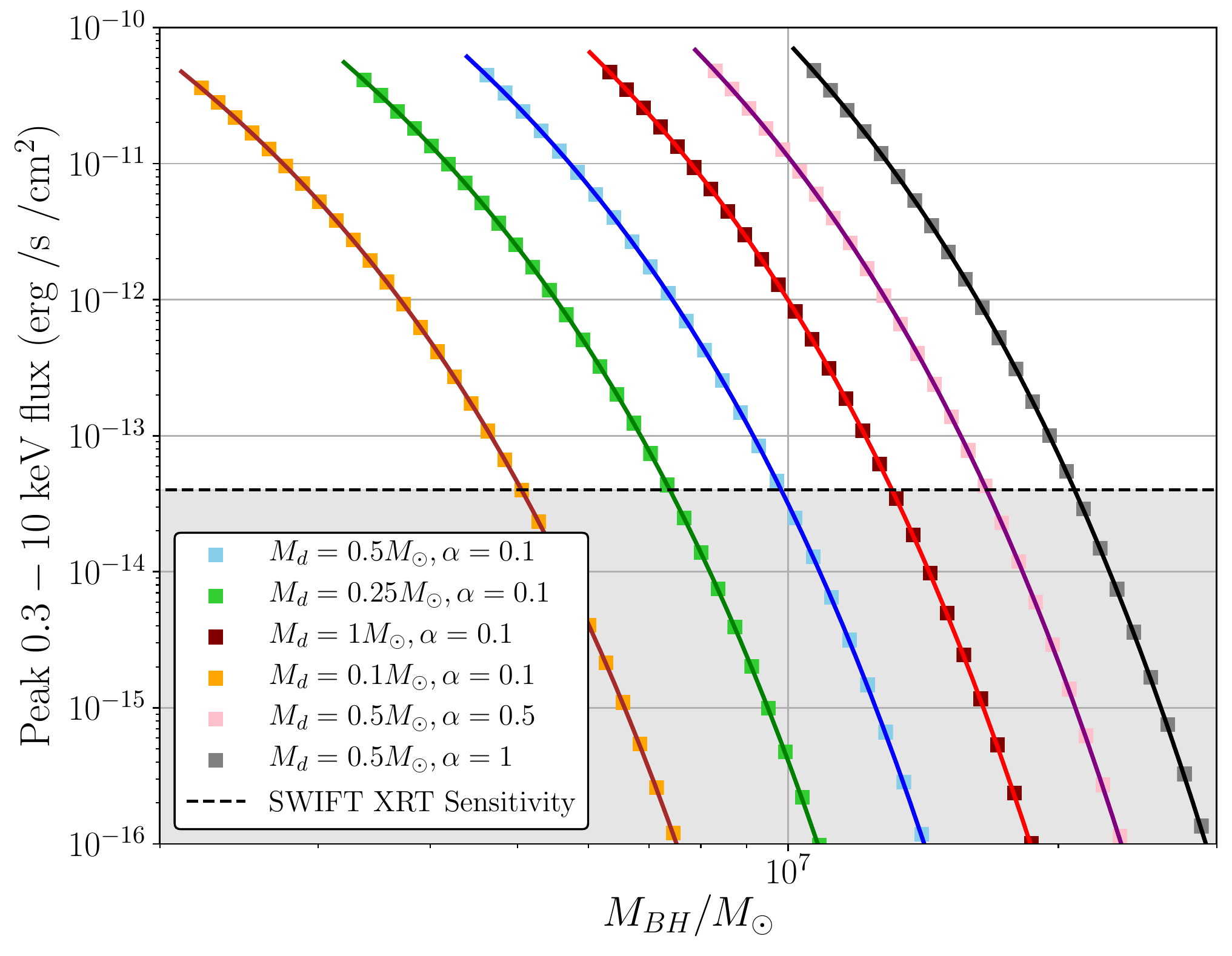} 
 \caption{The peak X-ray flux, as observed at $100$Mpc, for different  initial disc masses $M_d$ and $\alpha$ parameters as shown.  } 
 \label{edlim2}
\end{figure}

Once $F_0, \psi_1, \psi_2$ and $A_1$ have been determined  from a single set of numerically determined fluxes, the peak X-ray flux for different values of $\alpha, M_d$ and $M$ follows simply.  Figure (\ref{edlim2}) shows the peak observed X-ray fluxes as a function of black hole mass, for light curves with different disc parameters, as shown.   Here, we are interested only in parameter regimes where $L_{\rm bol} < L_{\rm Edd}$.   Since $L_{\rm bol} \propto M^{-8/3}$ (eq.\ \ref{bolop}), this will correspond to black hole masses in excess of a characteristic value we denote as $M_{\rm Edd}$.    This is a function of the disc mass and $\alpha$ parameter (eq. \ref{edratio}).  The solid curves in Figure \ref{edlim2} are the analytical model (eq. \ref{FX}).

 Figure \ref{edlim2} demonstrates that the functional form (eq. \ref{FX}) reproduces the numerical results very well for a wide range of physically reasonable disc parameters.  The exponential cut-off %{\bf (OK?)}
  of X-ray flux from large mass black holes means that TDE discs around black hole masses larger than a critical  value (denoted $M_{\rm lim}$), will be unobservable at X-ray energies.

\subsection{Summary}
\begin{figure}
  \includegraphics[width=.5\textwidth]{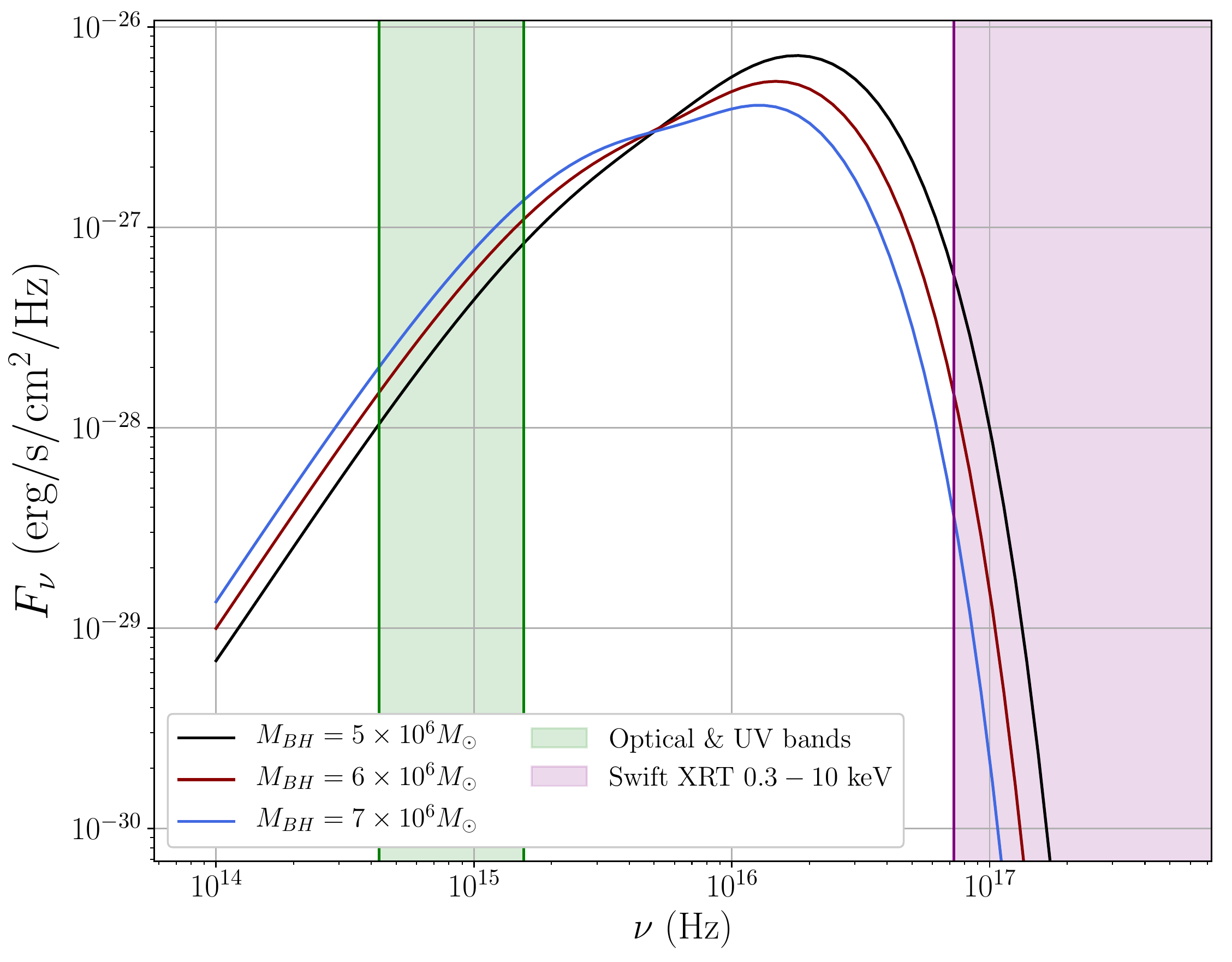} 
 \caption{The disc spectrum at the time of peak bolometric luminosity for three different discs, all with initial mass $M_d = 0.5 M_{\odot}$ and $\alpha=0.1$, for three different black hole masses. The disc spectra are calculated assuming $f_{\rm col} = 2$ when the disc temperature $T > T_{\star} = 7 \times 10^4$ K, and $f_{\rm col} = 1$ otherwise.  The green and purple shaded regions correspond to optical and {(part of the)} X-ray observing bands respectively.  Note that while the bolometric luminosities of these solutions are similar, the observable X-ray luminosity is strongly black hole mass-dependent.  } 
 \label{spectrum}
\end{figure}

We have presented numerical solutions of the relativistic thin disc equations, computing the peak observed $0.3 {\textendash} 10$ keV fluxes of discs evolving in a Schwarzschild spacetime for a wide variety of assumed disc parameters.   The assumed distance is 100Mpc, with a face-on disc orientation.   Our interest here is in sub-Eddington disc solutions.  For a given disc mass and $\alpha$ parameter, this constrains the black hole mass to be larger than a characteristic scale $M > M_{\rm Edd}$, the Eddington mass, which may be determined numerically. % {\bf (OK?)}.    
The X-ray flux is then exponentially cut-off as a function of black hole mass.   %{\bf (OK?)}   
The disc is unobservable at X-ray energies for black hole masses larger than a value $M_{\rm lim}$, the X-ray limiting mass.   

Note that while the X-ray luminosities of the disc solutions at the Eddington and X-ray black hole masses differ by a large amount ($L_{X, {\rm Edd}} / L_{X,{\rm lim}} \sim 1000$), the {\it bolometric} luminosities differ by a much smaller factor  ($L_{\rm bol,Edd}/L_{\rm bol, lim} = \left(M_{\rm lim}/M_{\rm Edd}\right)^{8/3} \sim 6$).   The total energetics from the bolometric light curves are therefore grossly similar;  it is only the respective spectral distributions which sharply differ.  This is especially clear in figure \ref{spectrum},  which shows a snapshot of the disc spectrum at the time at which the disc bolometric luminosities peak, for different black hole masses (denoted on the plot).  This demonstrates the heightened sensitivity of the observed X-ray flux to the peak disc temperature, in contrast to the much smaller variation of the bolometric disc luminosity for different black hole masses.  % {(\bf OK?)}  

\section{Limiting observable black hole mass}\label{imp}

\subsection{Simple physical picture}

As is clear from figure \ref{edlim2}, thin disc models of X-ray TDEs naturally lead to a maximum X-ray observable black hole mass for a given set of disc parameters.  % {\bf OK?)}   
While we have thus far analysed a somewhat restricted and simplified model, the sense of our results %{\bf (OK?)}  
is likely to be robust and to hold more generally.   The ISCO in higher mass black hole discs lies farther out in radius, where less energy is locally liberated  by the disc shear.   This naturally results in lower peak disc temperatures.   The X-ray luminosity is extremely sensitive to the temperature of the hottest part of the disc (eq. \ref{FX}), so the lower peak disc temperatures associated with larger mass black holes translates to much lower X-ray luminosities.

We present more general %{\bf (OK?)} 
detailed numerical calculation in sections (\S\ref{vansec} \&\ref{spinsec}).   Here, we focus on some important observational implications of the results at hand.   For example, even if a massive accretion disc is able to form efficiently in the aftermath of a TDE,  a high mass black hole will not result in an observable levels of X-rays.    By contrast, as is clear from fig \ref{spectrum}, the optical \& UV luminosities of the relativistic disc solutions are not sensitive to the black hole mass.    The dominant component of the optical \& UV luminosity from a TDE results from disc emission at late times, where the luminosity is observed to plateau (van Velzen {\it et al}. 2019, Mummery \& Balbus 2020a,b).   Our results therefore imply that extremely X-ray dim TDEs should still be observed to exhibit this characteristic UV-plateau at large times.  

\subsection{Mathematical description }
For the simplified numerical set-up of section \ref{num1}, the form of the dependence of the largest X-ray observable black hole %{\bf (OK?)}  
mass on disc parameters may be determined both analytically {(via equation \ref{maxmass_schwarz})}, as well as by direct  numerical calculation.  Numerically, we simply iterate computations of disc light curves until we find the black hole mass, for a given $M_d$ and $\alpha$, at which the observed flux equals  a prescribed limiting cut-off $f_{\rm lim}$. % {\bf (OK?)}   
For the {\it Swift}  X-ray telescope we use the value $f_{\rm lim} \equiv 4 \times10^{-14} \,{\rm erg/s/cm}^2$.  

\begin{figure}
\includegraphics[width=.5\textwidth]{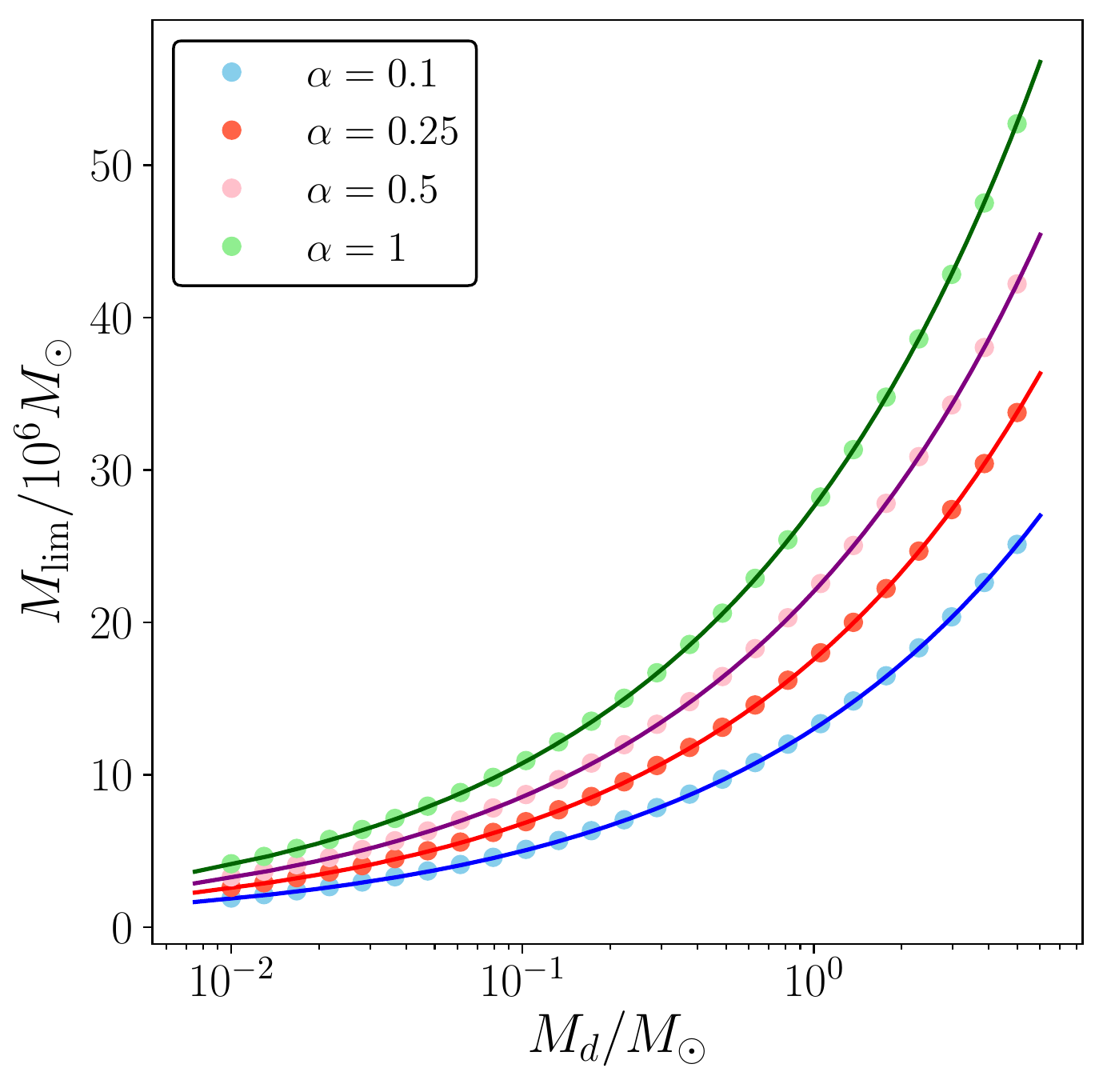} 
\caption{The maximum observable black hole mass, defined as the black hole mass at which the peak X-ray disc flux is  $4 \times10^{-14} \,{\rm erg/s/cm}^2$, as a function of initial disc mass, for four different $\alpha$ parameters at a distance of $100$ Mpc. The solid curves are the theoretical predictions of eq. \ref{maxmass_schwarz}, which reproduce the parameter dependence well.}
\label{massplot}
\end{figure}

To proceed analytically, we need to invert eq.\ [\ref{FX}] for the black hole mass
\beq
F_X(M,M_d,\alpha,D) = f_{\rm lim} \rightarrow {M_{\rm lim}} = {M_{\rm lim}}(\alpha, M_d,D) .
\eeq
We may derive a form for ${M_{\rm lim}}$ by retaining only the leading-order large black hole mass behaviour of the flux (equation \ref{full}).   
%In order to keep track of the dimensions of various quantities, 
We define dimensionless distance $d \equiv D/100$ Mpc, disc mass $m_d = M_d/0.5M_{\odot}$, and black hole mass $m = M/M_\star$, where $M_\star$ is defined so that $\Lambda = m^{7/6}/\alpha^{1/3}m_d^{5/12}$ {in eq. [\ref{lamb}].}     With an amplitude $f_0$ carrying the dimensions of flux, the equation we wish to invert has the form 
\beq\label{firstorderFX}
F_X \simeq { f_0 \over d^{2}} { \alpha^{2/3} m_d^{5/6} \over  m^{1/3}} \exp\left( - {m^{7/6} \over m_d^{5/12}  \alpha^{1/3}} \right) = f_{\rm lim} .
\eeq
Collecting terms and defining simplifying variables $x$ and $y$ by
\beq 
x \equiv \frac{m}{ \alpha^{2/7} \, m_d^{5/14}} ,
\eeq
and 
\beq
y \equiv \frac{f_{\rm lim}}{f_0} \frac{d^2}{ \alpha^{4/7} \, m_d^{5/7}} ,
\eeq
we are finally left with the compact equation 
\beq
y = x^{-1/3} \exp\left(-x^{7/6}\right) 
\eeq
to be inverted for $x(y)$.  
This may be solved easily when $x$ is small ($x\simeq 1/y^3$) or when $x$ is large  ($x\simeq|\ln y|^{6/7}$).  More generally, the inversion may be carried out by using the %real branch % {(\bf do we need 'real branch'?)} of the 
lambert $W$ function\footnote{The lambert $W$ function is widely available in, e.g. the Scipy  package.}, $W_0$ {(Corless {\it et al}. 1996)}:  
\beq
x = \left[ \frac{2}{7}\, W_0 \left(\frac{7}{2 y^{7/2}}\right)\right]^{6/7} ,
\eeq
and so 
\beq\label{maxmass_schwarz}
{M_{\rm lim}} = M_\star \,  \alpha^{2/7} m_d^{5/14} \left[ \frac{2}{7}\, W_0 \left(\frac{ 7\alpha^{2} \, m_d^{5/2}}{2d^7} \left(\frac{f_0}{f_{\rm lim}}\right)^{7/2} \right)\right]^{6/7}. 
\eeq
Figure \ref{massplot} shows the numerically computed maximum observable black hole masses, as a function of disc mass, for four different $\alpha$ parameters, with a set-up identical to that of section \ref{num1}.   Also plotted is the analytical expression (eq. \ref{maxmass_schwarz}; see Appendix \ref{param_vals} for numerical values of $f_0$ and $M_\star$).  % {\bf( Confusing, which parameters?)}.  
The analytical results reproduce the numerical results with great fidelity.

We see that  ``standard'' TDEs (those with disc masses $M_d < 1M_{\odot}$, and $\alpha \simeq 0.1$), observed nearly face-on around Schwarzschild black holes, should be X-ray observable only for $M \lesssim 10^7 M_{\odot}$. 

\section{Vanishing ISCO stress}\label{vansec}
TDEs represent a promising observational path to probe a contentious theoretical issue: does the dynamical disc stress vanish at the inner edge of a relativistic accretion disc?    Mummery \& Balbus (2019b) demonstrated that the properties of the ISCO stress directly control the discs temporal evolution, the bolometric luminosity from finite (vanishing) ISCO stress discs decreases more slowly (rapidly).  % {\bf(OK?)}   
The effects of the ISCO stress are particularly pronounced at X-ray energies, which generally originates in the near-ISCO region.   Detailed modelling of {\it individual} sources will always be the accurate probe of the properties of the ISCO stress.  For example, the TDE ASASSN-14li is best modelled as a disc with a finite ISCO stress (Mummery \& Balbus 2020a), whereas ASASSN-15lh is better modelled with a maximal spin and much smaller ISCO stress (Mummery \& Balbus 2020b).    Nevertheless, consideration of more general properties of the population of X-ray TDEs offers some valuable insight on the nature of the ISCO stress. 

In this section we present an overview of the steps required to perform the analogous \S \ref{deriv} calculation for the case in which the stress is assumed to vanish at the ISCO. 

\subsection{Analytical results}\label{vanderiv}

There is a key difference between the finite and vanishing ISCO stress disc models, which is particularly relevant when considering X-ray energies: in standard disc models,  the disc temperature vanishes at the ISCO.    This means that the temperature maximum occurs within the disc, and at a location defined by (see eq. \ref{kdef}), 
\beq
k_1 \equiv \left({R_p \over T_p}\right) \left.{{\partial T \over \partial R}}\right|_{R_p} = 0 .
\eeq
%{\bf (Check that everything here has been defined.)}
With $k_1$ vanishing, the leading order term relevant for the Laplace integral expansion is  the quadratic $k_2$ term (eq. \ref{betaexpansion}, \ref{kdef}).
This leads to a series involving gaussian integrals for the disc spectrum and X-ray flux, rather than exponential integrals.  Aside from this important detail, the procedure is identical to that performed in \S \ref{deriv}.   The gaussian form of the integrals changes the leading power law exponent in the X-ray flux expression, which now has the form 
\beq\label{FXVS}
F_X = F_0  \left(\frac{R_p}{D}\right)^{2}  \left[ \Lambda^{-3/2} + {\phi_1}\Lambda^{-5/2} + { \phi_2}\Lambda^{-7/2} \right] e^{-\Lambda} .
\eeq
In this expression, $\Lambda$ is defined in the same way as in section \ref{deriv} (eq. \ref{lamb}). The flux amplitude $F_0$ and the order-unity expansion coefficients $\phi_1$ and $\phi_2$ remain constants, but differ by order unity factors from the definitions used in section \ref{deriv}.  The disc temperature has the same  parameter dependence as in a finite ISCO stress disc, and so the leading order parameter dependence is given by 
\beq\label{FXVS_leadingorder}
F_X \propto   \frac{ \alpha^{1/2} M_d^{5/8}M^{1/4}}{D^2}  \exp\left(-A_1\frac{M^{7/6}}{ \alpha^{1/3} M_d^{5/12}}\right) .
\eeq
Similarly, a calculation of the upper observable mass limit for vanishing ISCO stress discs, analogous to that of \S \ref{imp},  can be performed.  Using the same  dimensionless variables as in \S\ref{imp}, eq. \ref{FXVS_leadingorder} now has the form 
\beq
F_X \simeq { f_0 \over d^{2}} \,{ \alpha^{1/2} m_d^{5/12}   m^{1/4}} \exp\left( - {m^{7/6} \over m_d^{5/12}  \alpha^{1/3}} \right) = f_{\rm lim}.
\eeq
This leads to 
\beq\label{maxvs}
{M_{\rm lim}} = M_\star \alpha^{2/7} m_d^{5/14} \left[ -{3\over 14} W_{-1} \left(-{14\over3}  y^{14/3} \right)\right]^{6/7},
\eeq
where
\beq
y \equiv {f_{\rm lim}\over  f_0 }{d^2 \over \alpha^{4/7} m_d^{5/7}},
\eeq
and $W_{-1}$ is the negative branch of the Lambert $W$ function.  All numerical values of relevant parameters are presented in Appendix \ref{param_vals}.   In the following section, we verify that these results reproduce the properties of the full numerical solutions of the relativistic disc equations. 

\subsection{Numerical results}\label{vannum}
\begin{figure}
  \includegraphics[width=.5\textwidth]{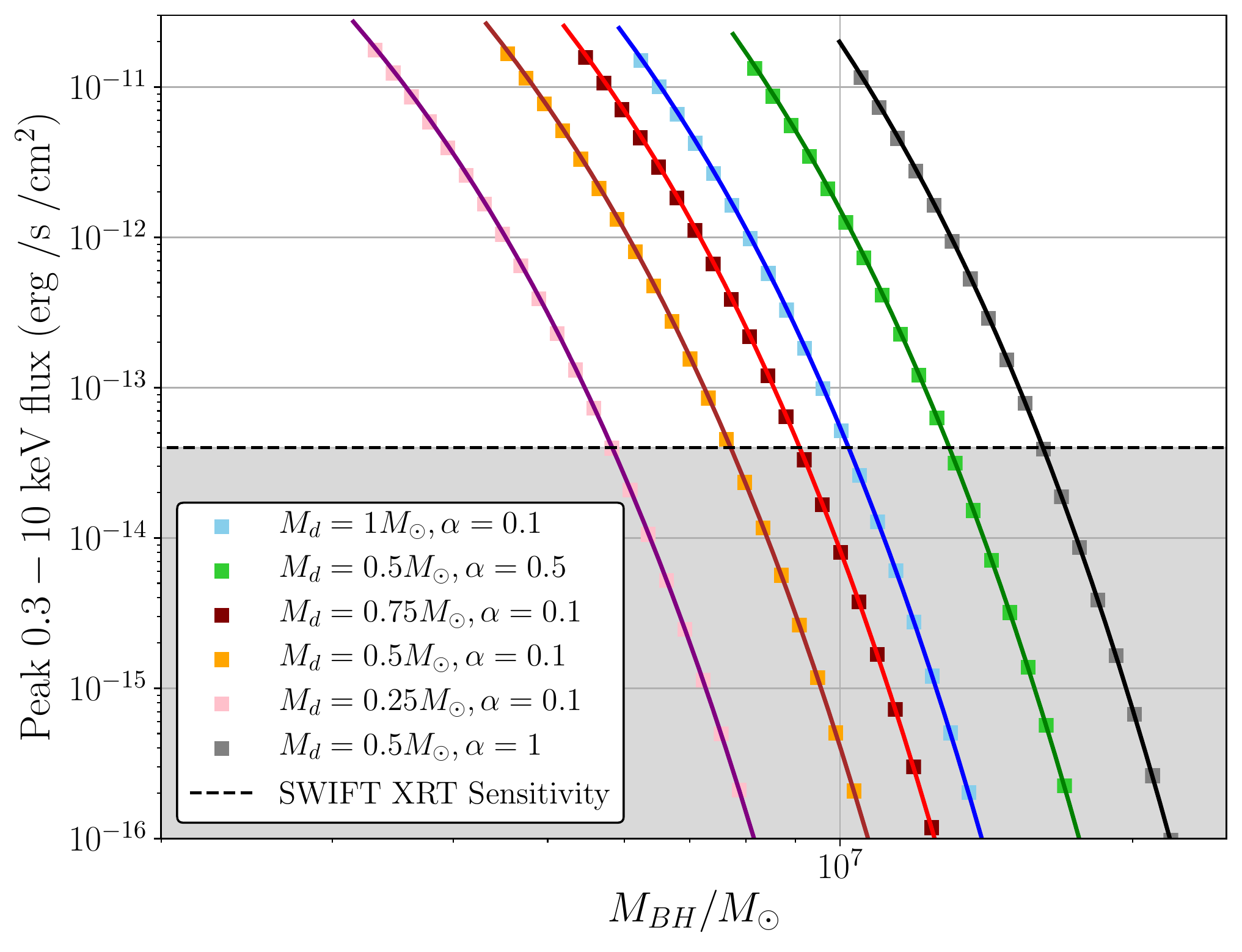} 
 \caption{The peak X-ray flux, as observed at $100$Mpc, for discs evolving with a number of different  initial disc masses $M_d$ and $\alpha$ parameters, denoted on plot. This plot was made with a vanishing ISCO stress.   } 
 \label{edlimVS}
\end{figure}

\begin{figure}
\includegraphics[width=.5\textwidth]{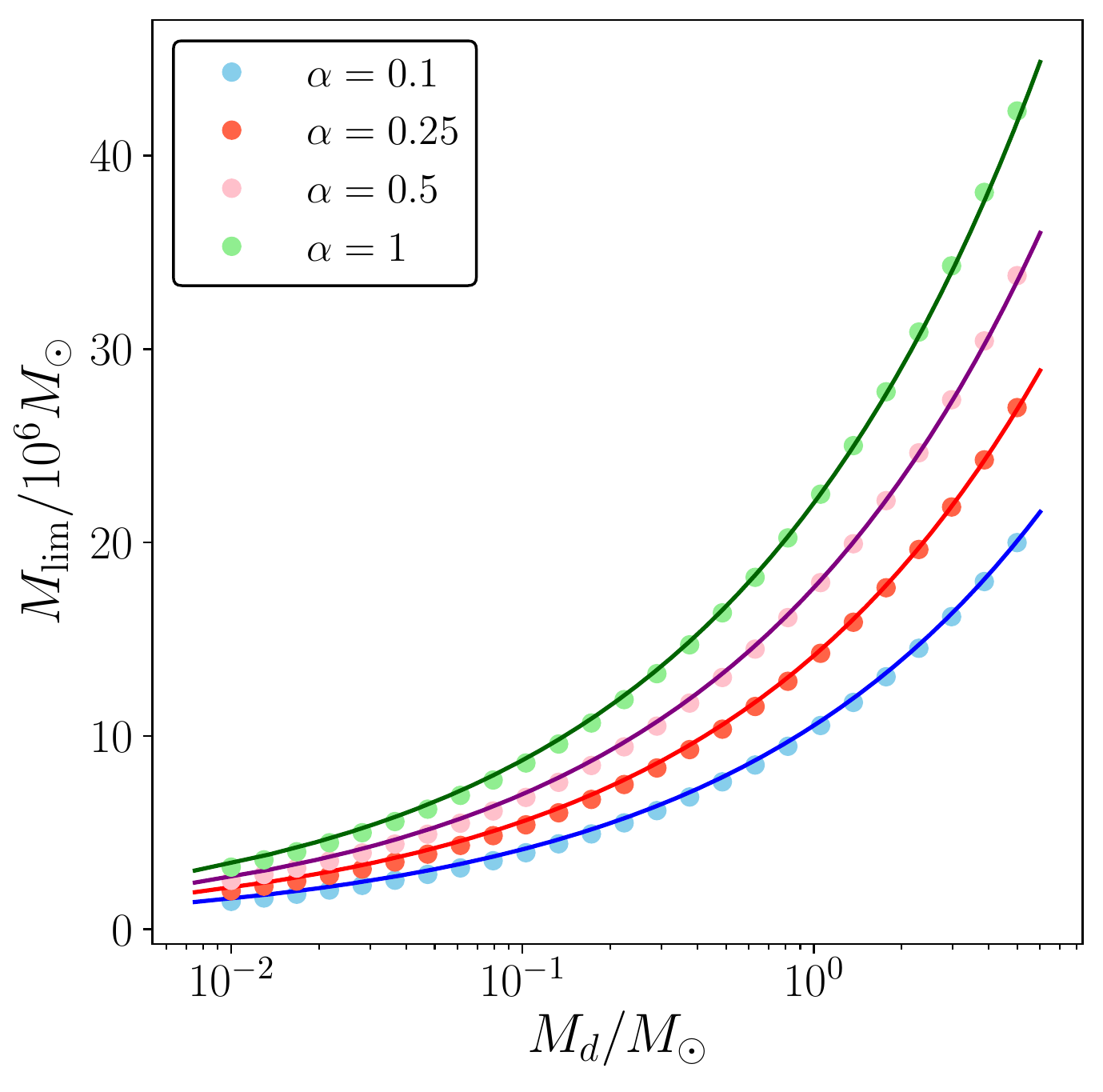} 
\caption{The maximum observable black hole mass, defined as the black hole mass at which the peak X-ray disc flux is  $4 \times10^{-14} \,{\rm erg/s/cm}^2$, as a function of initial disc mass for four different $\alpha$ parameters at a distance of $100$ Mpc. This plot was made with a vanishing ISCO stress. The solid curves are the theoretical predictions of eq. \ref{maxvs}, which capture the parameter dependence well.  } 
\label{massplotVS}
\end{figure}

In this section we perform a set of numerical experiments analogous to \S \ref{num1}, computing the peak value of the evolving X-ray light curves of accretion discs with various values of initial disc mass and $\alpha$ parameters for  $D = 100$ Mpc.  The same initial conditions are used; the only difference in the modelling is the ISCO stress value, which is set to zero throughout. 

Figure \ref{edlimVS} shows the peak observed X-ray fluxes as a function of black hole mass, for disc light curves using different disc parameters, as denoted on plot.  As before, we are interested only in the parameter regime where $L_{\rm bol} < L_{\rm Edd}$,  which corresponds to black holes more massive than $M_{\rm Edd}$.   Figure [\ref{edlimVS}] demonstrates that  eq.\ [\ref{FXVS}] reproduces the numerical results for a wide range of physically reasonable disc parameters.   Figure \ref{massplotVS} shows the maximum observable black hole masses {for a vanishing ISCO stress,} computed numerically with an identical set up as figure \ref{edlimVS}, as a function of disc mass for four different $\alpha$ parameters.  Also plotted is the analytical expression (eq. \ref{maxvs}), which reproduces the numerical results with great fidelity.  

Gross modo, the properties of the X-ray luminosity are relatively insensitive to the properties of the ISCO stress in the $L_{\rm bol} < L_{\rm Edd}$ regime.   The dominant behaviour in this regime for any ISCO stress is the diminution of X-rays arising from cooling discs of fixed $\alpha$ and $M_d$ about larger mass black holes, $T \propto M^{-7/6}$.   In quantitive detail, an otherwise identical discs (same $M$, $\alpha$, $M_d$, $f_{\rm col}$ etc.) with a finite ISCO stress would produce  a larger X-ray luminosity compared with their vanishing ISCO stress counterpart.  This results in finite ISCO stress discs being observable around black holes with somewhat larger masses.  It may well be the case that a population of X-ray bright, thermal TDEs around black holes of large masses ($M \gtrsim 2\times 10^7 M_\odot$) is indicative of non-zero ISCO stresses present in TDE discs. 

\section{Black hole spin and inclination angle}\label{spinsec}

The previous results have all been carried out for discs around Schwarzschild blackholes, observed face-on. Astrophysical black holes will generally have a non-zero angular momentum parameter, being described by the Kerr metric, and may of course  be observed at a general angle $\theta_{\rm obs}$.  In the following section, we revisit the numerical results of sections \ref{num1} \& \ref{vannum} for a variety of different blackhole spins and disc-observer inclination angles. 

In what follows, along with more general metrics and viewing angles, we shall also adopt the colour correction model of Done {\it et al}. (2012).  { This model is routinely used for the modelling of AGN disc spectra. It is likely that the disc conditions in TDEs will be most similar to those in AGN, and so this model should accurately model TDE disc colour-correction effects.}   For disc temperatures $T > 1\times10^5 {\rm K}$, electron scattering dominates the absorption opacity and the colour correction is assumed to saturate to
\beq\label{col1}
f_{\rm col}(T) =  \left(\frac{72\, {\rm keV}}{k_B T}\right)^{1/9}, \quad  T > 1\times10^5 {\rm K}. 
\eeq
For lower temperatures, the colour correction is an increasing function of temperature. This results from electron scattering no longer being the dominant source of absorption opacity, which is instead dominated by ionised Hydrogen and Helium.  Done {\it et al}. (2012) model the colour correction factor in this regime as 
\beq\label{col2}
f_{\rm col}(T) =  \left(\frac{T}{3\times10^4 {\rm K}} \right)^{0.83}, \, 3\times10^4 {\rm K} < T < 1\times10^5 {\rm K} .
\eeq
Note that the magnitude of $f_{\rm col}$ is continuous between the two regimes. Below the critical temperature $T = 3\times10^4 {\rm K}$ Hydrogen starts to become neutral and the associated Hydrogen absorption opacity becomes large. This results in the full thermalisation of the liberated disc energy, meaning the emitted disc spectrum is well described by a blackbody spectrum with
\beq\label{col3}
f_{\rm col}(T) = 1, \quad T < 3\times 10^4 {\rm K} . 
\eeq

\subsection{Schwarzschild black hole -- varying inclination}
\begin{figure}
  \includegraphics[width=.5\textwidth]{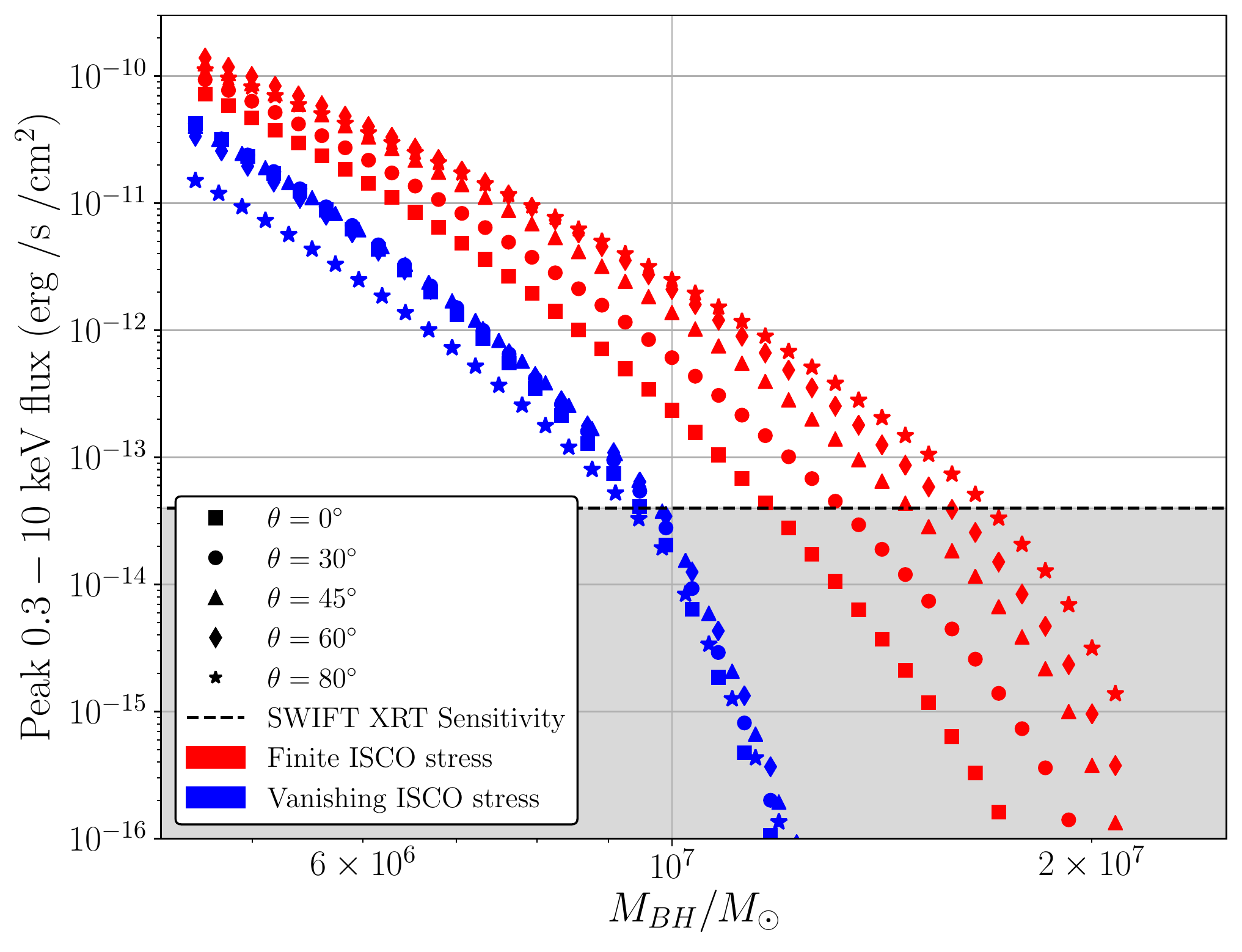} 
 \caption{The peak X-ray flux, as observed at $100$Mpc, for discs evolving with $M_d = 0.5 M_\odot$, $\alpha = 0.1$ for a number of different observer inclination angles, denoted on plot. Discs with a  vanishing ISCO stress are plotted in blue while discs with a finite ISCO stress are plotted in red. 
 } 
 \label{theta_schwarz}
\end{figure}
When a relativistic accretion disc is observed at an inclined angle, substantial Doppler blue shifting of photons emitted from the inner disc material moving with large line-of-sight velocities can overcome the gravitational red-shift.  This results in frequency ratio factors $f_\gamma > 1$, boosting the observed high-energy emission relative to a face-on orientation.  Counteracting this effect is the fact that disc regions moving away from the observer have large Doppler red shifts, and so the observed area of the hottest disc regions (those regions with maximum $\widetilde T = f_{\rm col} f_\gamma T$ ) decreases with increasing inclination. The net result of these two competing effects cannot be determined analytically and a detailed numerical calculations is required.  

For a Schwarzschild black hole the effects of inclination angle are modest, and the qualitative properties of the X-ray luminosity of the disc solutions at a general angle are quite similar to the face-on case.  Indeed, for a vanishing ISCO stress disc around a Schwarzschild black hole the inclination has almost no effect on the peak X-ray luminosity.    Finite ISCO stress discs are more sensitive to the inclination:  their temperature profiles peak closer to the central black hole, where changes in the Doppler boosting are more pronounced.   This is demonstrated in figure \ref{theta_schwarz}.  The enhanced Doppler boosting of the emitted radiation for more edge-on inclinations means that the maximum observable black hole mass is larger for larger observer inclinations.   The asymmetry of the photon red-shift factor $f_\gamma$ and the  disc temperature dependence of the colour correction factor (eqs. \ref{col1}, \ref{col2}, \ref{col3}) means that the analytical results derived in \S\ref{deriv} no longer describe the numerical results with great accuracy.    Their physical content is, however, very revealing of more general models.

\subsection{Rapid black hole spin -- varying orientation} 
\begin{figure}
  \includegraphics[width=.5\textwidth]{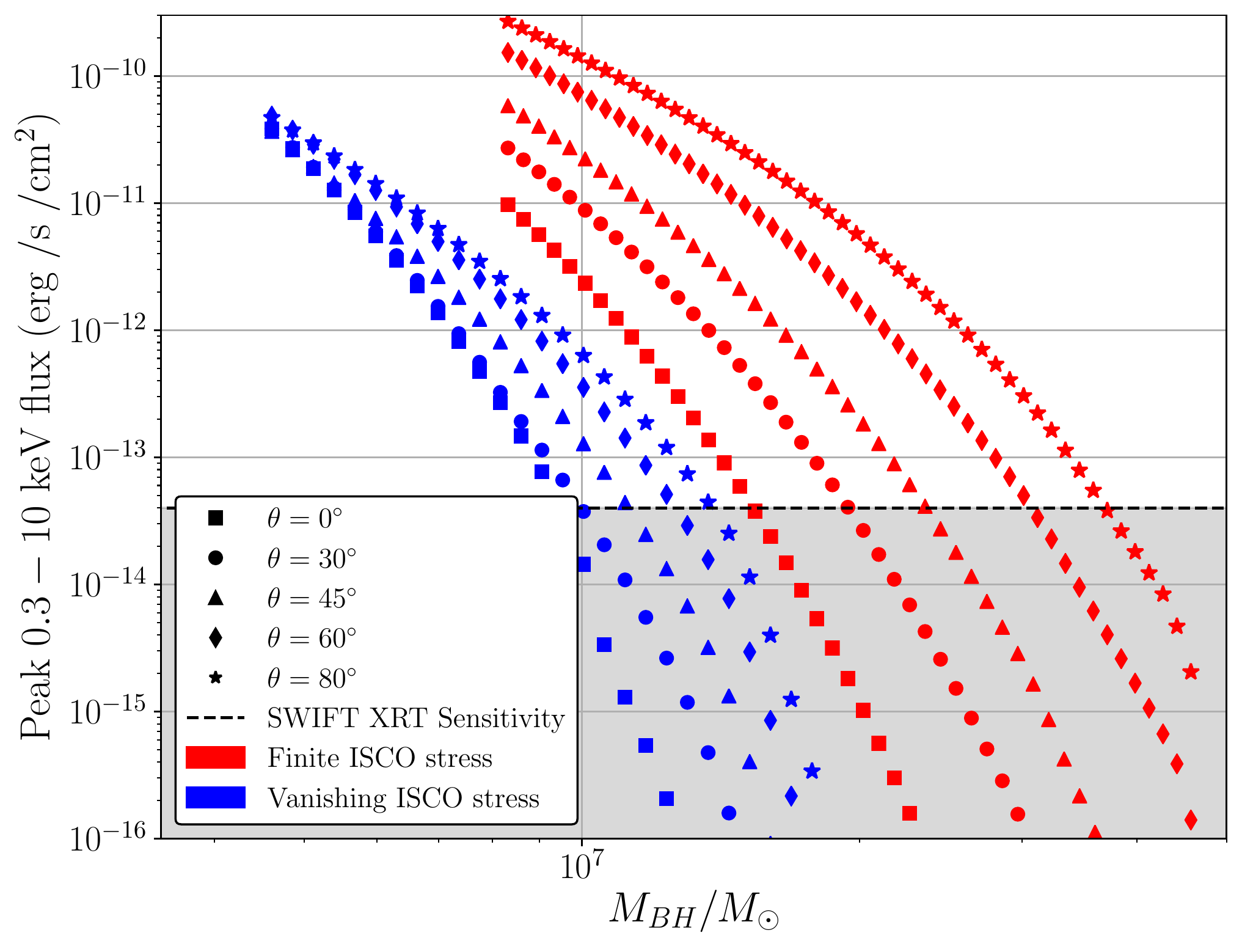} 
 \caption{The peak X-ray flux, as observed at $100$Mpc, for discs evolving with $M_d = 0.5 M_\odot$, $\alpha = 0.1$ for a number of different observer inclination angles, denoted on plot. The black hole spin is $a/r_g = 0.9$. Discs with a vanishing ISCO stress are plotted in blue while discs with a finite ISCO stress are plotted in red.   } 
 \label{theta_spin}
\end{figure}
Figure \ref{theta_spin} involves an identical set of numerical calculations as figure \ref{theta_schwarz}, except for the black hole spin, which corresponds to rapid rotation $a/r_g = 0.9$.   For higher spins, the differences between the different models of the ISCO stress are much more pronounced.  Like their steady state analogues, time dependent finite ISCO stress discs have intrinsically harder spectra, and larger mass-to-light efficiencies than vanishing ISCO stress discs (Agol \& Krolik 2000).  For larger black hole spins, these differences are more pronounced, and finite ISCO stress discs are much brighter at X-ray energies.  They can therefore be observed around black holes a factor of $\sim 2 $ more massive compared to vanishing ISCO stress discs.  

The finite ISCO stress solutions presented here were produced with a large ISCO stress.  In earlier work, we have argued that the general properties of numerical disc solutions should be thought of as part of a continuum between heavily stressed and vanishing stress solutions, controlled by the $\gamma$ parameter (Mummery \& Balbus 2019b), in effect a measure of the magnitude of the inner disc stress.  The current solutions should therefore be regarded as extremes:    a disc with a smaller, but non-zero, ISCO stress will lie somewhere between the two sets of results. 

Quantitively, the differences between the results of a rapidly spinning black hole and those of a Schwarzschild black hole are relatively modest, with only factor $\sim 1.5$ change in the maximum observable mass.  It is therefore a robust prediction that TDEs with thermal X-ray spectra will not be observed around black holes more massive than $M_{\rm lim} \sim 3 \times 10^7 M_{\odot}$.  The exceptions would arise for TDEs involving extremely massive stars or for discs with very large $\alpha$ parameters, $\alpha \simeq 1$. 

\section{The thermal X-ray TDE population}\label{compsec}

We next compare the properties of the black hole masses of the current thermal X-ray TDE population with the results of this paper.   The authors are aware of 12 X-ray TDEs with X-ray spectra which are well-modelled by thermal disc emission, and which have published estimates of the central black hole mass.  In view of the results presented here, we would predict that the black hole mass distribution should peak below $M_p < 10^7 M_\odot$, and that sources with masses $M \gtrsim 3\times 10^7 M_\odot$ should be rare.  With a sufficiently large disc mass thermal X-ray TDEs could be observed around black holes with extremely large masses, but the sparsity of high-mass stars (${\rm d}N_\star / {\rm d} M_\star \propto M_\star^{-2.35}$) should make these events uncommon. 

For the analysis, we use well-established galactic scaling relationships between the black hole mass and (i) the galactic bulge mass $M : M_{\rm bulge}$, (ii) the galactic velocity dispersion $M : \sigma$, and (iii) the bulge V-band luminosity $M : L_V$. All of the scaling relationships are taken from McConnell \& Ma (2013).  Where available, values of $ M_{\rm bulge}$, $\sigma$ and $L_V$ were taken from the literature for each TDE, and are presented in Table \ref{fulldatatable} in Appendix \ref{data}. The mean black hole mass for each TDE is then presented in Table \ref{datatable}. 

\begin{figure}
  \includegraphics[width=.5\textwidth]{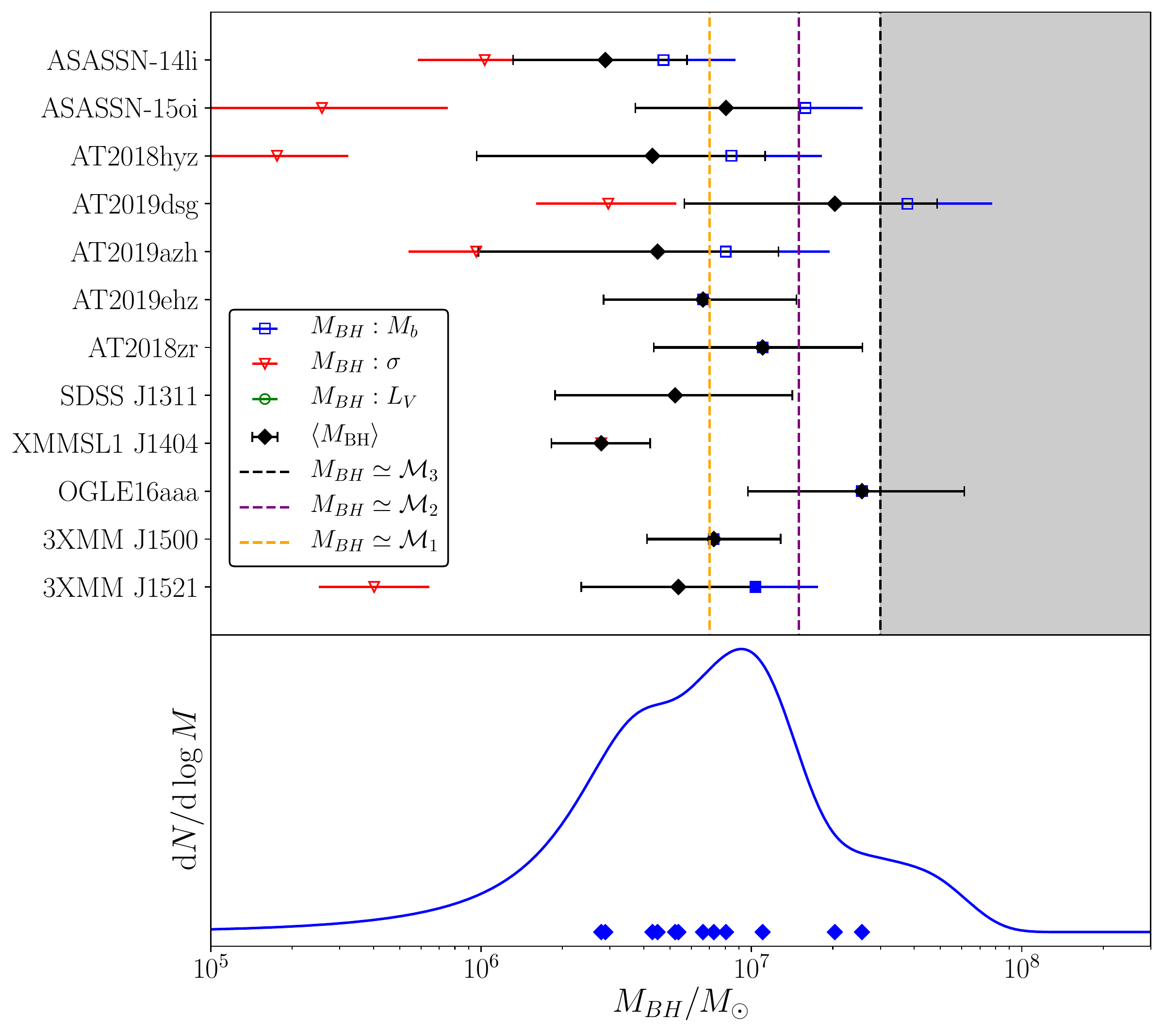} 
 \caption{The black hole mass distribution of the 12 TDEs with bright quasi-thermal X-ray spectra. The upper panel shows the black hole mass values inferred for each individual TDE, calculated using correlations with  (i) the galactic bulge mass $M : M_{\rm bulge}$ (blue squares), (ii) the galactic velocity dispersion $M : \sigma$ (red triangles), and (iii) the bulge V-band luminosity $M : L_V$ (green circles). The mean black hole mass for each TDE is plotted as a black diamond.  The lower panel shows the  black hole mass distribution of the quasi-thermal X-ray TDE population, obtained using kernel density estimation using a kernel width equal to the uncertainty in each TDEs black hole mass.  } 
 \label{thermpop}
\end{figure}

In the upper section of figure \ref{thermpop} we show the black hole masses of the twelve thermal X-ray TDEs obtained from each galactic scaling relationship, along with the mean black hole mass of each TDE (black diamond). We also display, as vertical dashed lines, three characteristic upper-observable black hole masses, corresponding to discs with masses $M_d = 0.05, 0.2$ and $0.5 M_\odot$,  and $\alpha$-parameter of $\alpha = 0.1$ about black holes of spin $a/r_g = 0.9$ at a distance of $D = 100$ Mpc and inclination angle $\theta = 60^\circ$. In the lower section of figure \ref{thermpop} we show the current distribution of the black hole masses of the thermal X-ray TDE population, obtained using kernel density estimation using a kernel width equal to the uncertainty in each TDEs black hole mass. 

\begin{table}
\renewcommand{\arraystretch}{2}
\centering
\begin{tabular}{|p{2.2cm}|p{2.cm}|p{2cm} |}\hline
TDE name  & $\left\langle M_{\rm BH}\right\rangle/10^6M_\odot$ \\ \hline\hline
ASASSN-14li & $2.9^{+2.9}_{-1.6} $  \\ \hline
ASASSN-15oi & $8.1^{+7.1}_{-4.3}$ \\ \hline
AT2018hyz & $4.3^{+6.9}_{-3.3} $ \\ \hline
AT2019dsg & $20.4^{+28.3}_{-14.7} $ \\ \hline
AT2019azh & $4.5^{+8.0}_{-3.5} $ \\ \hline
AT2019ehz &$6.6^{+8.0}_{-3.8}$ \\ \hline
AT2018zr & $11.0^{+14}_{-6.7}  $ \\ \hline
SDSS J1311 & $5.2^{+8.9}_{-3.3} $ \\ \hline
XMMSL1 J1404 ${}^a$ & $2.8^{+1.4}_{-1.0} $ \\ \hline
OGLE 16aaa & $26.0^{+35}_{-16} $ \\ \hline
3XMM J1521 & $5.4^{+5.1}_{-3.0}$ \\ \hline
3XMM J1500 & $7.3^{+5.5}_{-3.2}$ \\ \hline
\end{tabular}
\caption{The mean black hole mass of the 12 Thermal X-ray TDEs from the literature. ${}^a$ The TDE XMMSL1 J1404 also has non-thermal X-ray components present, but the non-thermal component is subdominant (Saxton et al. in prep, Wevers 2020).  }
\label{datatable}
\end{table}

It is clear that the current thermal X-ray TDE population is consistent with the expected results of this paper, the distribution peaks at a black hole mass $M \simeq 9 \times 10^6 M_\odot$, and is strongly suppressed above $M \gtrsim 3 \times 10^7M_\odot$.  It is interesting that one of the TDEs with the largest inferred black hole mass,  
AT2019dsg had jetted radio emission associated with it (Stein {\it et al}. 2020).  In our model, bright { thermal} X-ray emission from black holes this large should result via the disruption of an unusually large star by a black hole with large black hole spin,  precisely the sort of physical scenario  in which a jet may be launched. 

{ An interesting black hole mass scale with which to compare our results to is the so-called Hills mass (Hills 1975). The Hills mass $M_H$ is defined as the black hole mass at which a solar type star would be swallowed whole by the black hole before it reaches its tidal radius and undergoes tidal destruction. A simple estimate (using Newtonian gravity) gives $M_H \simeq 9\times 10^7 M_\odot$, although this mass scale increases by an order of magnitude for extreme values of the black hole spin parameter (Kesden 2012), and generally increases with stellar mass. Given that this mass scale will set an upper limit on the black hole mass distribution of {\it all} types of TDEs, it is important to demonstrate that the suppression of bright thermal X-ray TDEs above $M \sim 10^7 M_\odot$ is a result of the mechanism set out in this paper, and not merely this Hills mass effect.  

A simple way to test whether the observed suppression of high black hole mass TDEs observed with thermal X-ray spectra is a result of the mechanism set out in this paper is to ask how many TDEs of other spectral types have been observed with black hole mass $M > 3\times 10^7 M_\odot$. In companion papers to this work (Mummery \& Balbus 2021b, Mummery 2021a) we examine TDEs from across the entire black hole mass range. We find twelve TDEs with inferred black hole masses greater than $10^7 M_\odot$, and six TDEs with inferred black hole masses $M > 3\times 10^7 M_\odot$. None of these six highest black hole mass TDEs were observed to have thermal X-ray spectra. 

}

We stress that the argument here is not that TDEs observed at X-ray energies around black holes of masses $M > 10^7 M_\odot$  are themselves intrinsically rare.  { There are in fact as many of these sources (nine) as their are X-ray TDEs with masses less than $10^7M_\odot$.}  
{ The mass distribution of the total X-ray TDE population is approximately flat from $M \sim 10^6 - 10^8 M_\odot$ (Wevers {\it et al}. 2019).  }   The point is that the dominant emission components from these TDEs with larger masses are observed to be { dominated by} nonthermal { components} (Wevers 2020), not disc-like { thermal components}. 

{ The black hole mass dependence of the dominant emission components of different X-ray TDEs can be understood within the framework developed here, and is the focus of a companion paper (Mummery \& Balbus 2021b). In brief, if TDEs behave like scaled up analogues of the X-ray binaries observed in our own galaxy, then we would expect a growing nonthermal X-ray component to dominate at lower disc  Eddington ratios. Nonthermal emission, resulting from the Compton up-scattering of soft disc photons from an electron scattering corona, is expected to dominate when the accretion disc changes state at $l = L_{\rm bol}/L_{\rm edd} \sim 0.01$ (e.g. Fender \& Belloni 2004). Given the strong dependence of disc Eddington ratio on black hole mass $l \propto M^{-11/3}$ (eq. \ref{edratio}), we would expect this nonthermal component to dominate in TDEs of the largest black hole masses $M \gtrsim 3\times 10^7 M_\odot$. This nonthermal component will allow TDEs around large black hole masses, which would be unobservable with pure thermal X-ray emission, to be observable in the X-ray band. 

While TDEs evolving in the harder accretion state will produce observable levels of nonthermal emission around the largest black hoe mass TDEs, it is important to recognise that, even in those TDEs with nonthermal/coronal emission components present, the thermal components of the X-ray flux of these sources will still be well described by the scaling relationships developed in this paper. The upper black hole mass limit of $M \sim 3 \times 10^7 M_\odot$ for thermal-dominated X-ray TDEs is a robust prediction, and will not be modified by including small nonthermal components. 
}In the coming years,  wide-field X-ray surveys are expected to discover many more thermal X-ray TDEs, { and will rigorously test these predictions}.   We would expect the discovered populations to follow the qualitative distribution predicted by this paper, namely, that the majority { of bright thermal X-ray TDEs} will occur around { black holes} with $M \simeq 5 \times 10^6 M_\odot$, with very few occurring around high-mass ($M > 3\times10^7M_\odot$) black holes.

\section{Wide field surveys with low cadence}\label{survey}
\begin{figure}
  \includegraphics[width=.5\textwidth]{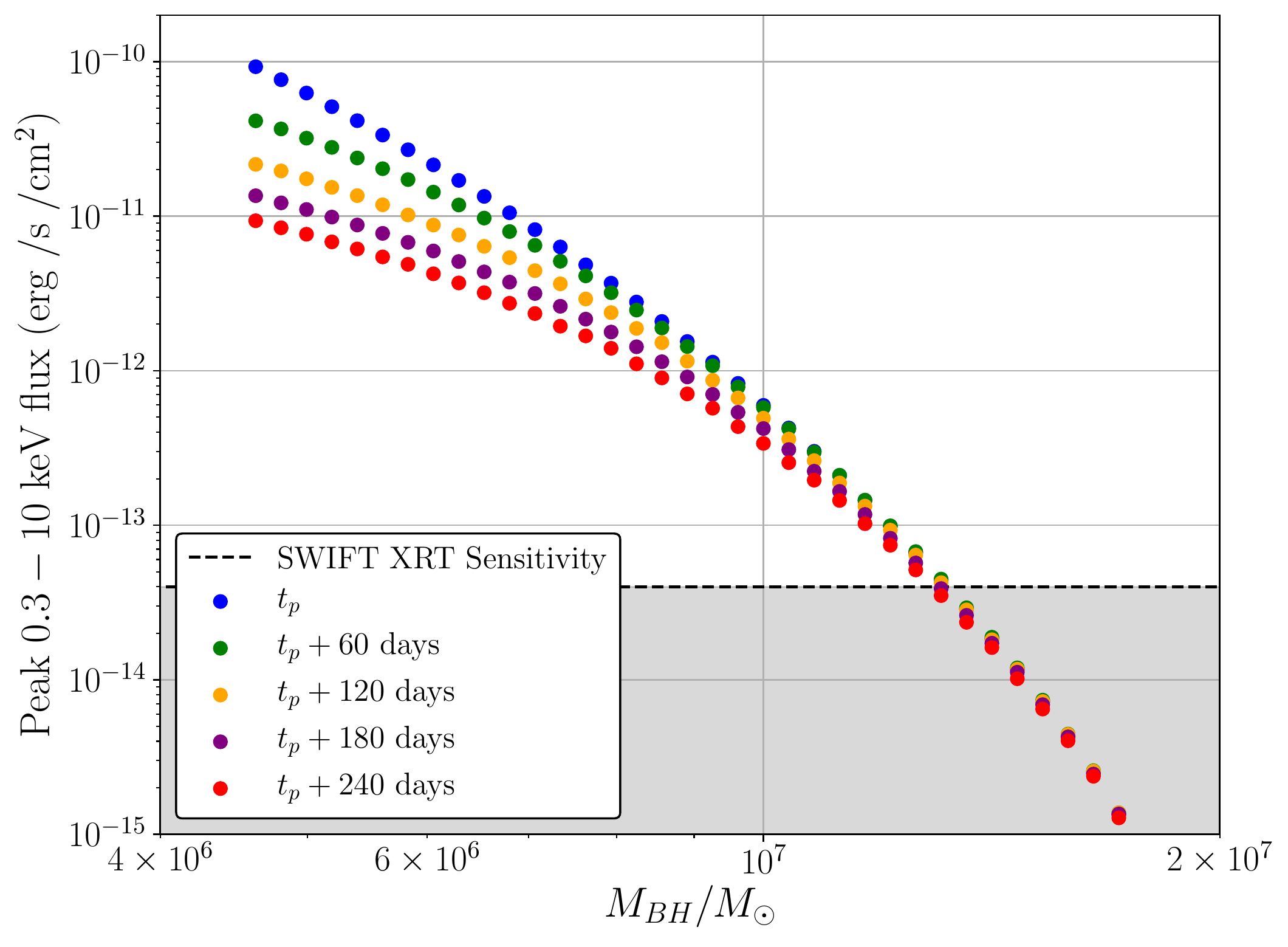} 
 \caption{The peak X-ray flux, as observed at $100$Mpc, for discs evolving with $M_d = 0.5 M_\odot$ and $\alpha = 0.1$ at a number of different times post peak, denoted on plot. The black hole spin is $a = 0$, a finite ISCO stress was assumed and the inclination angle is $\theta_{\rm obs} = 30^\circ$. } 
 \label{Fxtime}
\end{figure}
Near-future wide-field X-ray surveys are predicted to expand the sample of X-ray TDEs by one or two orders of magnitude.   For example, the Einstein Probe is expected to find $\sim$ 100 new TDEs per year (Yuan et al. 2015), while eROSITA is expected to find $\sim$1000 (Khabibullin et al. 2014).  While these surveys are expected to discover many TDEs, they have extremely low cadence (every patch of sky is observed once every 6 months by eROSITA, for example).  The quantitive implications of the sensitive parameter dependence of the disc X-ray flux on system parameters (particularly the black hole mass) for the observed TDE rates (and other observed properties) of these up-coming surveys lies beyond the scope of the current work.  
We shall, however, briefly consider how this low cadence could affect the implied detectable mass limits presented here. 

Once its maximum value has been reached, 
the temperature of the hottest point within a relativistic accretion disc is given approximately by (Mummery \& Balbus 2020a)
\beq
T_p(t) \simeq T_\star \left({t\over t_{\rm visc}}\right)^{-n/4}, \quad t > t_{\rm peak}, 
\eeq
where $n$ is the bolometric luminosity decay index, $L \sim t^{-n}$, which ranges from $n \simeq 0.5 - 1.2$ depending on the properties of the ISCO stress (Mummery \& Balbus 2019b). 
This time dependence, when combined with the exponential dependence on peak temperature of the X-ray flux (eq. \ref{FX}) results in faster-than-power-law ($\sim \exp[-t^{n/4}]$) decays in the observed X-ray flux from disc-dominated TDE sources.   (This was observed in the source ASASSN-14li [Mummery \& Balbus 2020a].)  The dependence of the viscous timescale on system parameters is clearly import for determining for how long a TDE accretion disc is observable at X-ray energies.   The viscous timescale generally scales like (Balbus \& Mummery 2018)% {\bf REFERERENCE.}

\beq
t_{\rm visc} \propto {\sqrt{GM R_0^3} \over \W} ,
\eeq
which for our stress parameterisation (eq. \ref{alphastress}) becomes 
\beq\label{tvisc}
t_{\rm visc} \propto {M^{8/3} \over \alpha^{4/3} M_d^{2/3}} .
\eeq
As can be seen by combining equations \ref{bolop} \&  \ref{tvisc}, this result means that the combination $L_{\rm bol, peak} \, t_{\rm visc}$, a proxy for the radiated energy, depends only upon the disc mass, as it must. 
Although they are generally much dimmer, thermal X-ray TDEs around more massive black holes will evolve much more slowly than those around lower mass black holes. This is demonstrated in figure \ref{Fxtime}. Although discs with larger $\alpha$ parameters do evolve significantly more quickly, the estimates of the maximum observable black hole masses derived in this paper are not significantly affected, even for surveys with low cadence like eROSITA. 

\section{Conclusions}\label{conc}
In this paper we have modelled, analytically and numerically, the properties of thermal X-ray emission emergent from relativistic time-dependent accretion discs, tailoring our analysis to parameter regimes  most appropriate for comparison with TDEs. Our key result demonstrates a strong suppression of thermal X-ray emission from accretion discs around black holes with large masses (eq. \ref{full}). 

This strong X-ray suppression for TDEs around larger mass black holes results in a maximum observable black hole mass for thermal X-ray TDEs, with thermal emission only observable around black holes with masses $M \lesssim 3\times10^7M_\odot$. Both the properties of the X-ray luminosity and upper observable black hole mass limit are a function of disc parameters, and the full dependence can be described analytically (eq. \ref{FX} \& eq. \ref{maxmass_schwarz}). We have demonstrated that the current population of observed X-ray TDEs is indeed consistent with an upper black hole mass limit of order $M \sim 10^7M_\odot$, consistent with our analysis (Figure \ref{thermpop}). Our results make quantitive predictions about the distribution of TDEs discovered by upcoming wide-field X-ray surveys, and will be directly tested in the coming years.

\section*{Data accessibility statement}
All data used in this paper is presented in full in Appendix \ref{data}. 

\section*{Acknowledgements} 
This work is partially supported by STFC grant ST/S000488/1, and the Hintze Family Charitable Foundation. It is a pleasure to acknowledge useful conversations with Jeremy Goodman, and very constructive comments from our referee.

\appendix{}
\section{Different opacity parameterisations}\label{opacity_params}
The exact parameterisation of the opacity within the disc slightly modifies the dependence of the peak disc temperature on the black hole mass, disc mass, and $\alpha$ parameter. In this Appendix we calculated the power-law dependence of the disc surface temperature on system parameters, assuming a general bi-power-law disc opacity relationship: 
\beq\label{opac}
\kappa = \kappa_0 \rho^a T_c^{-b} ,
\eeq
which may be used for a wide range of physically plausible models, including electron scattering $a = b = 0$, or a Kramers opacity: $a = 1, b = 7/2$. 
To aid readability we reproduce the governing equations from the main text (section \ref{scalings}):
\begin{align}
\W &\equiv \alpha r c_s^2,  \label{Astressdef} \\ 
c_s^2 &= {P_g \over \rho} =  \frac{k_B T_c}{\mu m_p} , \label{Acs} \\
T_c^4 &= \frac{3}{8} \kappa \Sigma T^4 ,\label{ATc} \\
\sigma T^4 &= -\frac12 \W \Sigma \Omega' . \label{AT}
\end{align}
Three more relationships are required to close the set of equations for the general opacity law \ref{opac}. The density of the disc is trivially related to the disc surface density and scale height $H$ through:
\beq\label{rhodef}
\rho = \Sigma/H.
\eeq
This scale height $H$ is then related to the orbital frequency $\Omega$ and sound speed $c_s$ by:
\beq\label{Hdef}
c_s^2 = H^2 \Omega^2 ,
\eeq
finally, the orbital frequency $\Omega$ is, in the Newtonian limit, given by
\beq\label{finaleq}
\Omega = \sqrt{{GM \over r^3}}.
\eeq
The set of equation \ref{opac}--\ref{finaleq} suffice to fully determine the parameter dependence of the disc surface temperature. The substitution of equation \ref{Astressdef} \& \ref{AT} into equation \ref{ATc}, followed by simplifications using equations \ref{opac}, \ref{Acs}, \ref{rhodef} \& \ref{Hdef} leads to 
\beq
T_c^{3 + b + a/2} = - {3 \over 16 \sigma} \kappa_0  \left({k_B \over \mu m_p}\right)^{1-a/2} \alpha r  \Sigma^{2+a} \Omega^a  \Omega' .
\eeq 
Equation \ref{finaleq} implies that $\Omega' = -3\Omega/2r$, so that the central temperature of the disc is given by
\beq
T_c = \left[ {9 \over 32 \sigma} \kappa_0 \left({k_B \over \mu m_p}\right)^{1-a/2}  \alpha \Sigma^{2+a} \Omega^{1+a}\right]^{2/(6+2b+a)}.
\eeq
The equations \ref{Astressdef}, \ref{Acs} \& \ref{AT} together demonstrate that the surface temperature of the disc scales like
\beq
T^4 \propto \alpha T_c \Sigma \Omega 
\eeq
or in terms of the variables $\alpha, \Sigma$ \& $\Omega$:
\beq
T^4 \propto \alpha^{1 + Q} \, \Sigma^{1 + (2+a)Q} \, \Omega^{1 + (1 +a)Q} ,
\eeq
where $Q \equiv 2/(6+2b+a)$. Finally, using the general scaling relationships $\Omega \propto 1/M$, $\Sigma \propto M_d/M^2$, we are left with 
\beq
T^4 \propto \alpha^{X} M_d^{Y} M^{Z}
\eeq
with indices
\begin{align}
X &= (8 + 2b + a)/(6 + 2b + a) , \\
Y &= (10 + 2b + 3a)/(6 + 2b + a) , \\
Z &= - (28 + 6b + 9a)/(6 + 2b + a) .\label{massdepgen}
\end{align}
The  system parameter dependence of the surface temperature is therefore only weakly dependent on the exact opacity specification. As reported in section \ref{scalings} an electron scattering opacity results in
\beq
T_p \propto {\alpha^{1/3} \, M_d^{5/12} \over M^{7/6}}, \quad a = b = 0.
\eeq
whereas Kramers opacity would result in 
\beq
T_p \propto {\alpha^{2/7} \, M_d^{5/14} \over M^{29/28}}, \quad a = 1,  b = 7/2.
\eeq
The key results in this paper result from the pronounced decrease in peak disc temperature as the central black hole mass is increased. Equation \ref{massdepgen} demonstrates that this is a general property of these thin disc solutions, and is not dependent on a particular stress or opacity parameterisation. 

\section{Numerical values of fitting parameters}\label{param_vals}
In this Appendix we provide the numerical values of the fitting parameters $F_0, M_\star, \psi_1$ \& $\psi_2$. 
These parameters are required to analytically compute the X-ray flux and upper observable black hole mass scale. 
\subsection{Finite ISCO stress}
The analytical expression for the X-ray flux (eq. \ref{FX}) has 4 free parameters, and can be written in the following form
\beq\label{BFX}
F_X = F_0 \left(\frac{m}{d}\right)^{2}  \left[ \frac{1}{\Lambda^2} + \frac{\psi_1}{\Lambda^3} + \frac{ \psi_2}{\Lambda^4} \right] e^{-\Lambda} .
\eeq
We have defined a dimensionless black hole mass $m \equiv M/M_\star$, and source-observer distance $d \equiv D/100$ Mpc. A final free parameter sets the amplitude of the important $\Lambda$ parameter 
\beq
\Lambda \equiv {h\nu_l \over k_B \T_p} = A_1 {M^{7/6} \over \alpha^{1/3} M_d^{5/12} } .
\eeq
If we define $m_d \equiv M_d / 0.5 M_\odot$, then $\Lambda$ may be written
\beq
\Lambda \equiv { m^{7/6} \over \alpha^{1/3} m_d^{5/12} } .
\eeq
It is this expression that determines the magnitude of the fitted parameter $M_\star$. The remaining fitted parameters are then found using the numerically calculated fluxes of Fig. \ref{naivemass}. The best fitting parameters are:
\begin{align}
F_0 &= 8.07 \times 10^{-9} \, {\rm erg/s/cm}^2, \\
M_\star &= 2.50 \times 10^6 M_\odot, \\
\psi_1 &= 5.5, \\
\psi_2 &= 7.0.
\end{align}
The flux amplitude required to fit the numerically calculated upper-observable black hole mass scales (equation \ref{maxmass_schwarz}) differs slightly from $F_0$. This is a result of dropping the $\psi_1$ \& $\psi_2$ correction terms in the derivation of equation \ref{maxmass_schwarz}. The best-fitting value $f_0$ used to calculate Fig. \ref{massplot} is
\beq
f_0 = 1.21 \times 10^{-8}  \, {\rm erg/s/cm}^2 .
\eeq
\subsection{Vanishing ISCO stress}
The analytical expression for the vanishing ISCO stress X-ray flux (eq. \ref{FXVS}) similarly has 4 free parameters, and can be written in the following form
\beq\label{BFXVS}
F_X = F_0  \left(\frac{m}{d}\right)^{2}  \left[ \Lambda^{-3/2} + {\phi_1}\Lambda^{-5/2} + { \phi_2}\Lambda^{-7/2} \right] e^{-\Lambda} .
\eeq
An identical procedure as in the proceeding section (i.e, defining $\Lambda \equiv {\alpha^{1/3} m_d^{5/12} / m^{7/6}}$, before fitting to numerically calculated fluxes) leads to 
\begin{align}
F_0 &= 2.17 \times 10^{-8} \, {\rm erg/s/cm}^2, \\
M_\star &= 1.67 \times 10^6 M_\odot, \\
\phi_1 &= 5.5, \\
\phi_2 &= 7.0.
\end{align}
The flux amplitude relevant for calculating the upper observable black hole mass (eq. \ref{maxvs}) is
\beq
f_0 = 3.26 \times 10^{-8}  \, {\rm erg/s/cm}^2 .
\eeq

\section{TDE Black hole masses from galactic scaling relationships}\label{data}
To analyse  the current black hole mass distribution of thermal X-ray TDEs we use well-established galactic scaling relationships between the black hole mass and (i) the galactic bulge mass $M : M_{\rm bulge}$, (ii) the galactic velocity dispersion $M : \sigma$, and (iii) the bulge V-band luminosity $M : L_V$. All of the scaling relationships are taken from McConnell \& Ma (2013).  Where available, values of $ M_{\rm bulge}$, $\sigma$ and $L_V$ were taken from the literature for each TDE, and are presented in Table \ref{fulldatatable}.

\begin{table*}
\renewcommand{\arraystretch}{2}
\centering
\begin{tabular}{|p{2.2cm}|p{2.cm}| p{2. cm} | p{1.5 cm} | p{2. cm} | p{2. cm} | p{2. cm} | p{1.2 cm} |}
\hline
TDE name  & $M_{\rm bulge}/M_\odot$  &  $M_{\rm BH}/M_\odot$ & $\sigma$ (km/s)  &  $M_{\rm BH}/M_\odot$ & $L_{V}/L_\odot$  &  $M_{\rm BH}/M_\odot$ &References \\ \hline\hline
ASASSN-14li & $2.0  \times 10^{9}$ & $4.7^{+4.0}_{-2.2} \times 10^{6}$ & $78 \pm 2$ & $1.0^{+0.79}_{-0.45} \times 10^{6}$ & --- & --- &[1], [2] \\  \hline
ASASSN-15oi & $6.3 \times 10^{9}$ & $1.6^{+1.0}_{-0.61} \times 10^{7}$ & $61 \pm 7$ & $2.6^{+5.0}_{-1.7} \times 10^{5}$  & --- & --- & [3],[4] \\  \hline
AT2018hyz & $3.5^{+0.9}_{-0.7} \times 10^{9}$ & $8.4^{+9.8}_{-4.7} \times 10^{6}$ & $57 \pm 1$ & $1.8^{+1.4}_{-0.8} \times 10^{5}$ & --- & --- &[5],[6] \\  \hline
AT2019dsg & $1.4^{+0.6}_{-0.4} \times 10^{10}$ & $3.8^{+4.0}_{-2.1} \times 10^{7}$ & $94 \pm 1$ & $3.0^{+1.5}_{-1.1} \times 10^{6}$ & --- & --- &[5],[7] \\  \hline
AT2019azh & $3.3^{+1.3}_{-1.0} \times 10^{9}$ & $8.0^{+11.4}_{-5.0} \times 10^{6}$ & $77 \pm 2$ & $9.5^{+7.2}_{-4.1} \times 10^{5}$ & --- & --- &[5],[4] \\  \hline
AT2019ehz & $2.7^{+0.7}_{-0.56} \times 10^{9}$ & $6.6^{+8.0}_{-3.8} \times 10^{6}$ & --- & --- & --- & --- &[5] \\  \hline
AT2018zr & $4.5^{+1.8}_{-1.3} \times 10^{9}$ & $1.1^{+1.4}_{-0.67} \times 10^{7}$ & --- & --- & --- & --- &[8] \\  \hline
SDSS J1311 & --- & --- & --- & --- & $5.5 \pm 0.5 \times 10^8$ & $5.2^{+8.9}_{-3.3} \times 10^6$ & [9] \\  \hline
XMMSL1 J1404 ${}^a$  & ---  & --- & $93 \pm 1$ &$2.8^{+1.4}_{-1.0} \times 10^{6}$ & --- & --- & [10] \\  \hline
OGLE 16aaa & $1.0^{+0.58}_{-0.37} \times 10^{10}$ & $2.6^{+3.5}_{-1.6} \times 10^{7}$ & --- & --- & --- & --- & [11] \\  \hline
3XMM J1521 & $4.2 \times 10^9$ & $1.0^{+0.8}_{-0.4} \times 10^7$ & $66$ & $4.0^{+2.4}_{-1.5} \times 10^5$ & --- & --- &[12]  \\\hline
3XMM J1500 & $3 \times 10^9$ & $7.3^{+5.5}_{-4.2} \times 10^6$ & --- & --- & --- & --- &[13]  \\
\hline
\end{tabular}
\caption{The properties of the central black hole of the 12 Thermal X-ray TDEs form the literature. [1] Holoein {\it et al}. (2016a), [2] Wevers {\it et al}. (2017), [3] Holoein {\it et al}. (2016b), [4] Wevers {\it et al}. (2019), [5] van Velzen {\it et al}. (2020), [6] Short {\it et al}. (2020), [7] Cannizzaro {\it et al}. (2020), [8] van Velzen {\it et al}. (2019), [9] Maksym {\it et al}. (2010), [10] Wevers (2020), [11] Wyrzykowski {\it et al}. (2017), [12] Lin {\it et al.} (2015), [13] Lin {\it et al}. 2017.  ${}^a$ The TDE XMMSL1 J1404 also has non-thermal X-ray components present, but the non-thermal component is subdominant (Saxton {\it et al}. in prep, Wevers 2020). { The blackhole masses in the third, fifth and seventh columns correspond to the blackhole masses calculated using the galactic measurement in the column directly to their left. }  }
\label{fulldatatable}
\end{table*}

For some TDE hosts only the total galactic mass was available, rather than the bulge mass. In these cases we assume a fixed fraction ($50\%$) of the host mass is in the bulge.  The vast majority of the uncertainty in each inferred black hole mass measurement results from intrinsic scatter in galactic scaling relationships. Unfortunately, some host measurements have no reported uncertainties. In these cases the entirety of the black hole mass uncertainty results from intrinsic scatter in the scaling relationships. 

When a TDE has multiple independent black hole mass estimates we calculate the mean black hole mass $\left\langle M \right\rangle = {1\over N} \sum_{i=1}^N M_i$. The (asymmetric) uncertainty on this  mean mass is taken to be
\beq\label{error_bar_def}
\left\langle \sigma^\pm \right\rangle = \left({1\over N}{\sum_{i=1}^{N} (\sigma^\pm_i)^2}\right)^{1/2} ,
\eeq
where $\sigma^+/\sigma^-$ correspond to the upper/lower uncertainties respectively. 
The mean black hole masses of each TDE candidate are displayed in Table \ref{datatable}.

{ Unfortunately, some of the sources examined here have black hole mass estimates from different scaling relationships which are all formally precise (in the sense that they have small error bars), while being mutually inconsistent (often by over an order of magnitude) with other estimates for the blackhole mass of the same TDE. In these cases equation \ref{error_bar_def} leads to error ranges which do not encompass all of the different black hole mass estimates computed from different scaling relationships. To counter this problem we replaced the definition (eq. \ref{error_bar_def}) with the simple range of the multiple measurements, however this had no effect on the results of the source analysis (Fig. \ref{thermpop}). As the exact treatment of the black hole mass uncertainty did not effect the results of the analysis, we treated every source in an identical manner using equation \ref{error_bar_def}. 

  }

\label{lastpage}

 %%%%%%%% %%%%%%%% %%%%%%%% %%%%%%%% %%%%%%%% %%%%%%%%

\begin{thebibliography}{99}
\bibitem{}
Agol E., Krolik J. H., 2000, ApJ, 528, 161
\bibitem{}
Balbus, S.\ A.,\ 2014, MNRAS, 444, L54
\bibitem { }
Balbus, S.\ A.\ 2017, MNRAS, 471, 4832
\bibitem{}
Balbus S. A., \& Mummery A., 2018, MNRAS, 481, 3348
\bibitem{}
Bardeen J. M., Press W. H., Teukolsky S. A., 1972, ApJ, 178, 347
\bibitem{}
Bender C., Orszag S., 1978, Advanced Mathematical Methods for Scientists
and Engineers. McGraw-Hill, New York
\bibitem{}
Bright, S. J., et al., 2018, MNRAS,  475,  3,  4011
\bibitem{}
Cannizzaro G., Wevers T., Jonker P.~G., P{\'e}rez-Torres M.~A., Moldon J., Mata-S{\'a}nchez D., Leloudas G., et al., 2020, arXiv, arXiv:2012.10195
\bibitem{}
Corless, R.M., Gonnet, G.H., Hare, D.E.G. et al. 1996,. Adv Comput Math 5, 329
\bibitem{}
Davis S.~W., Done C., Blaes O.~M., 2006, ApJ, 647, 525
\bibitem { }
Done C., Davis S.~W., Jin C., Blaes O., Ward M., 2012, MNRAS, 420, 1848
\bibitem {}
Done, C., Gierli\'nski, M., \& Kubota, A.\ 2007, A\&ARv, 15, 1
\bibitem { }
Eardley, D.\ M., \& Lightman, A.\ P.\ 1975, ApJ, 200, 187
\bibitem{}
Fender R., Belloni T., 2004, ARA\&A, 42, 317
\bibitem{}
Fragile, P.\ C., Etheridge, S.\ M., Anninos, P., Mishra, B., \& Klu\'zniak, W.\ 2018, ApJ, 857, 1
\bibitem{}
Gezari S., Cenko S.~B., Arcavi I., 2017, ApJL, 851, L47
\bibitem{}
Hills J. G., 1975, Nature, 254, 295
\bibitem{}
Holoien T. W.-S. et al., 2016a, MNRAS, 455, 2918
\bibitem{}
Holoien T. W.-S. et al., 2016b, MNRAS, 463, 3813
\bibitem{}
Jiang, Y.\ F., Stone, J.\ M., \& Davis, S.\ W.\ 2013, ApJ 778, 65
\bibitem{}
Jonker P.~G., Stone N.~C., Generozov A., van Velzen S., Metzger B., 2020, ApJ, 889, 166
\bibitem{}
Kesden M., 2012, Phys. Rev. D, 85, 024037
\bibitem{}
Khabibullin I., Sazonov S., Sunyaev R., 2014, MNRAS, 437, 327
\bibitem{}
Li, L.-X., Zimmerman, E. R., Narayan, R., \& McClintock, J. E. 2005, ApJ, 157, 335 
\bibitem { }
Lightman A.\ P., Eardley D. M., 1974, ApJ, 187, L1
\bibitem{}
Lin D., Maksym P.~W., Irwin J.~A., Komossa S., Webb N.~A., Godet O., Barret D., et al., 2015, ApJ, 811, 43
\bibitem{}
Lin D., Guillochon J., Komossa S., Ramirez-Ruiz E., Irwin J.~A., Maksym W.~P., Grupe D., et al., 2017, NatAs, 1, 0033
\bibitem { }
Lynden-Bell, D., \& Pringle, J.\ E.\ 1974, MNRAS, 168, 603\ 
\bibitem{}
Maksym W.~P., Ulmer M.~P., Eracleous M., 2010, ApJ, 722, 1035
\bibitem{}
McConnell N.~J., Ma C.-P., 2013, ApJ, 764, 184
\bibitem{}
Miller J. M., et al., 2015, Nature, 526, 542
\bibitem{}
Misner, C. W., Thorne, K. S., \& Wheeler, J. A. 1973, Gravitation (San Francisco:
Freeman)
\bibitem{}
Mummery, A.,  \& Balbus, S.\ A., 2019a, MNRAS, 489, 132 
\bibitem{}
Mummery, A.,  \& Balbus, S.\ A., 2019b, MNRAS, 489, 143  
\bibitem { }
Mummery, A.,  \& Balbus, S.\ A., 2020a, MNRAS, 492, 5655
\bibitem{}
Mummery, A., \& Balbus, S.~A., 2020b, MNRAS, 497, L13
\bibitem{}
Mummery, A., \& Balbus, S.~A., 2021b, MNRAS, 504, 4730
\bibitem{}
Mummery, A., 2021a, MNRAS, 504, 5144
\bibitem{}
Mummery, A., 2021b, arXiv, arXiv:2104.06212
\bibitem{}
 Saxton R.~D., Read A.~M., Esquej P., Komossa S., Dougherty S., Rodriguez-Pascual P., Barrado D., 2012, A\&A, 541, A106
\bibitem { }
Shakura, N.\ I., \& Sunyaev, R.\ 1973, AA, 24, 337
\bibitem{}
Shimura T., Takahara F., 1995, ApJ, 445, 780
\bibitem{}
Short P., et al., 2020, arXiv, arXiv:2003.05470
\bibitem{}
Stein R., van Velzen S., Kowalski M., Franckowiak A., Gezari S., Miller-Jones J.~C.~A., Frederick S., et al., 2020, arXiv, arXiv:2005.05340
\bibitem {}
van Velzen, S., Stone, N. C., Metzger, B. D., Gezari, S., Brown, T. M. \& Fruchter, A. S.\  2019, ApJ, 878, 82 
\bibitem{}
van Velzen S., Gezari S., Hammerstein E., Roth N., Frederick S., Ward C., Hung T., et al., 2020, arXiv, arXiv:2001.01409
\bibitem{}
Wen S., Jonker P.~G., Stone N.~C., Zabludoff A.~I., Psaltis D., 2020, ApJ, 897, 80
\bibitem{}
Wevers, T., et al., 2017, MNRAS, 471, 1694 
\bibitem{}
Wevers T., Stone N.~C., van Velzen S., Jonker P.~G., Hung T., Auchettl K., Gezari S., et al., 2019a, MNRAS, 487, 4136
\bibitem{}
Wevers T., Pasham D.~R., van Velzen S., Leloudas G., Schulze S., Miller-Jones J.~C.~A., Jonker P.~G., et al., 2019b, MNRAS, 488, 4816
\bibitem{}
Wevers T., 2020, MNRAS, 497, L1
\bibitem{}
Wyrzykowski {\L}., Zieli{\'n}ski M., Kostrzewa-Rutkowska Z., Hamanowicz A., Jonker P.~G., Arcavi I., Guillochon J., et al., 2017, MNRAS, 465, L114
\bibitem{}
Yuan W., Zhang C., Feng H., Zhang S.~N., Ling Z.~X., Zhao D., Deng J., et al., 2015, arXiv, arXiv:1506.07735
\end{thebibliography}
\end{document}